\documentclass[pra,a4paper,twocolumn,superscriptaddress,longbibliography]{revtex4-1}
\usepackage{amssymb}
\usepackage{amsmath}
\usepackage{amsfonts}
\usepackage{graphicx}
\usepackage{bm}
\usepackage{xcolor}
\usepackage{multirow}

\DeclareMathOperator{\sgn}{sgn}

\DeclareMathOperator{\Si}{Si}

\begin{document}

\title{Interaction of a N\'eel--type skyrmion and a superconducting vortex}

\author{E. S. Andriyakhina}

\affiliation{Moscow Institute for Physics and Technology, 141700 Moscow, Russia}

\affiliation{\hbox{L.~D.~Landau Institute for Theoretical Physics, acad. Semenova av. 1-a, 142432 Chernogolovka, Russia}}

    \author{I. S. Burmistrov}

\affiliation{\hbox{L.~D.~Landau Institute for Theoretical Physics, acad. Semenova av. 1-a, 142432 Chernogolovka, Russia}}

\affiliation{Laboratory for Condensed Matter Physics, HSE University, 101000 Moscow, Russia
}

\date{\today} 

\begin{abstract}
Superconductor--ferromagnet heterostructures hosting vortices and skyrmions are new area of an interplay between superconductivity and magnetism. We study an interaction of a N\'eel--type skyrmion and a Pearl vortex in thin heterostructures due to stray fields. Surprisingly, we find that it can be energetically favorable for the Pearl vortex to be situated at some nonzero distance from the center of the N\'eel--type skyrmion. The presence of a vortex--antivortex pair is found to result in increase of the skyrmion radius. Our theory predicts that a spontaneous generation of a vortex--anti-vortex pair is possible under some conditions in the presence of a N\'eel--type skyrmion.  
\end{abstract}

\maketitle

\section{Introduction}

Topological objects have been remaining at the focus of theoretical and experimental research for more than half a century. The existence of topologically stable configurations in ferromagnets with Dzyaloshinskii--Moriya interaction has been predicted by Bogdanov and Yablonskii \cite{Bogdanov1989}. Now these topological excitations, termed as skyrmions, are intensively explored in an emergent field of {\it skyrmionics}~\cite{Back2020}. 

Research on an interplay between magnetism and superconductivity in heterostructures has long history~\cite{Ryazanov2004,Lyuksyutov2005,Buzdin2005,Bergeret2005,Eschrig2015}. Recently superconductor--ferromagnet bilayers hosting skyrmions have attracted great theoretical interest. It was understood that skyrmions in proximity with a superconductor can not only induce Yu-Shiba-Rusinov-type bound states \cite{Pershoguba2016,Poyhonen2016}
but can also host Majorana modes \cite{Chen2015,Yang2016,Gungordu2018,Mascot2019,Rex2019,Garnier2019,Rex2020}. It was found \cite{Yokoyama2015} that the presence of skyrmions affects strongly Josephson current via superconductor--ferromagnet--superconductor junction.  It has been also shown \cite{Vadimov2018} that skyrmion configurations can be stabilized by a superconducting dot or antidot situated at the top of a ferromagnetic film. 
In ferromagnet--superconductor heterostructures superconducting  vortices and skyrmions can form bound pairs either due to interplay of proximity effect and spin-orbit coupling \cite{Hals2016,Baumard2019} or due to their interaction via stray fields \cite{Dahir2019,Menezes2019,Dahir2020,Petrovic2021}. 

\begin{figure}[b]
    \centering
    \includegraphics[width=0.47\textwidth]{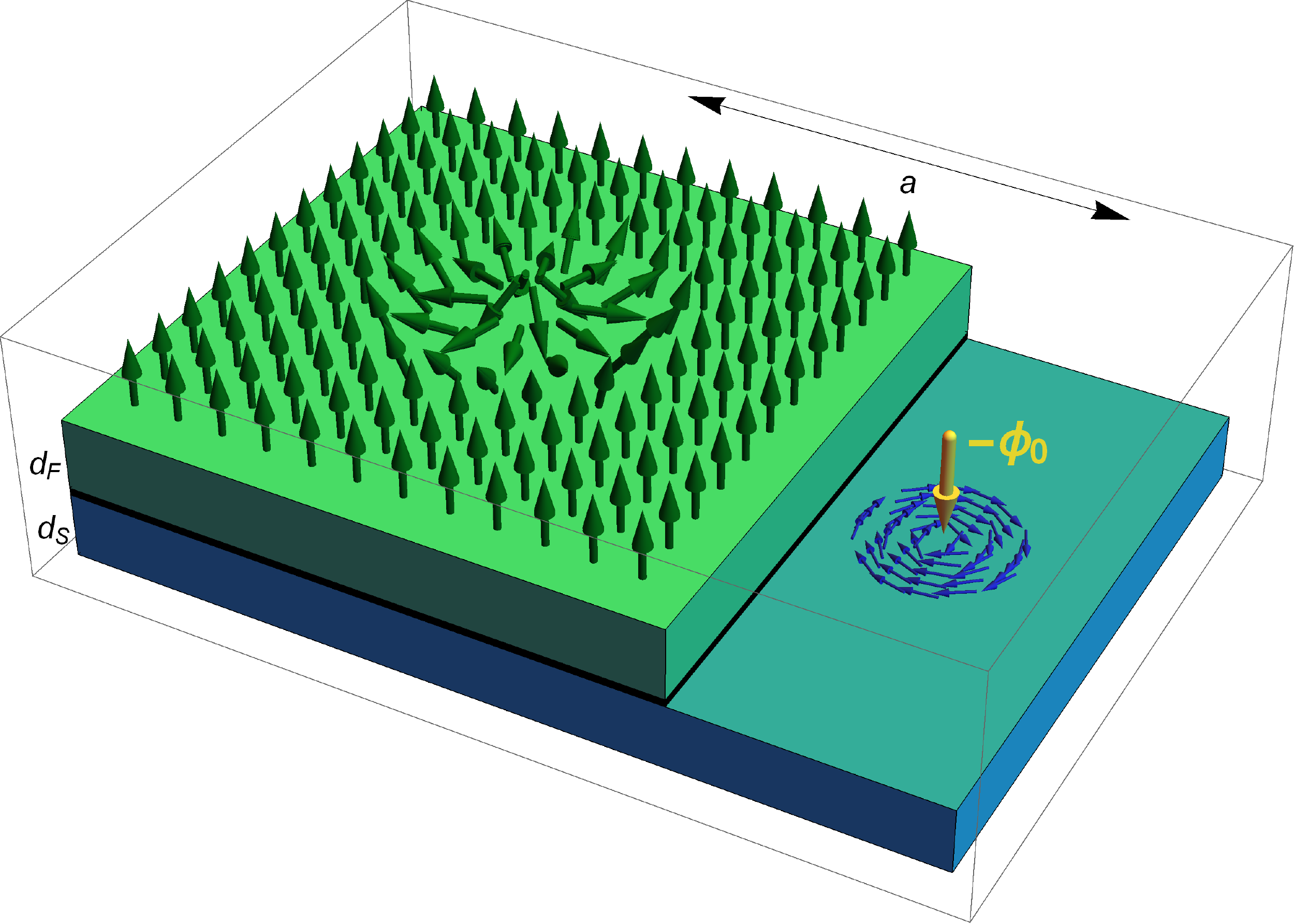}
    \caption{Sketch of a ferromagnet (green) -- superconductor (blue) heterostructure. There is also a thin insulating layer (black) which suppresses the proximity effect. The ferromagnetic layer hosts a N\'eel--type skyrmion. The magnetic profile of the skyrmion with the positive chirality is schematically shown. The superconducting layer hosts a vortex at some distance from the skyrmion's center. The vortex is shown schematically by blue lines,  the yellow arrow points towards the direction of its magnetic flux. $d_F$ and $d_S$ denote the width of the ferromagnet and superconductor film, respectively (see text).}
  \label{fig:Figure1}
\end{figure}

In this paper we study the interaction between a N\'eel--type skyrmion and a superconducting vortex in a chiral ferromagnet--superconductor heterostructure, see Fig. \ref{fig:Figure1}. We assume that the proximity effect is suppressed by the presence of a thin insulating layer between ferromagnet and superconductor such that the interaction between a skyrmion and a vortex is due to stray fields only. At first, by solving Maxwell--London equation we determine the Meissner current induced by a N\'eel--type skyrmion in the superconductor. Contrary to the previous work \cite{Dahir2020}, we consider the case of ferromagnet and superconducting films of arbitrary widths. Analysis of the general expression, cf. Eq. \eqref{eq:supercurrent}, in the case of thin ferromagnetic and superconducting films yields that the supercurrent has a maximum at distance of the order of the skyrmion size from the center of the skyrmion. Secondly, for thin ferromagnetic and superconducting films we compute the interaction energy between a N\'eel--type skyrmion  and a Pearl vortex due to stray fields.
Contrary to previous results, see Refs. \cite{Dahir2019,Menezes2019,Dahir2020}, we find that in the case of a N\'eel--type skyrmion with the positive and negative chiralities it can be energetically favorable for a vortex to settle at some distance from the skyrmion's center. At third, we study the effect of the presence of superconducting vortex--anti-vortex pair on the skyrmion size in thin heterostructures. We find that a Pearl  vortex leads to increase of a skyrmion radius. Under some conditions, the spontaneous generation of a vortex--anti-vortex pair in a superconducting film is possible in the presence of a skyrmion.

The outline of the paper is as follows. In Sec. \ref{Sec:SuperCurrent} the solution of the Maxwell--London equation is presented, and the results for the supercurrent are given. The interaction energy between a N\'eel--type skyrmion and a Pearl vortex is computed and analyzed in Sec. \ref{Sec:Interaction}. In Sec. \ref{Sec:BackAction} the effect of a Pearl vortex on the skyrmion radius is estimated. We end the paper with summary and conclusions in Sec. \ref{Sec:DiscConc}. Some technical details of computations are presented in Appendix.

\section{Supercurrent generated by a N\'eel--type skyrmion\label{Sec:SuperCurrent}}

We start from calculation of the supercurrent in the chiral ferromagnet--supercondutor heterostructure which is generated by a N\'eel--type skyrmion (see Fig. \ref{fig:Figure1}). The width of the chiral ferromagnet (superconductor) film is $d_F$ ($d_S$). We assume the presence of a thin insulating layer between the chiral ferromagnet and the superconductor  that allows us to neglect the proximity effect. The magnetization profile of a N\'eel--type skyrmion in the chiral ferromagnet film in the cylindrical coordinate system with the origin at the center of the skyrmion is given as follows \cite{Kawaguchi2016}
\begin{equation}
    \bm{M}_{\rm Sk} = M_s 
    \Bigl[ \bm{e_r} \eta \sin \theta (r)  + \bm{e_z} \cos \theta (r) \Bigr ] .
    \label{eq1}
\end{equation}
Here $\eta = \pm 1$ denotes the chirality of the skyrmion, $\theta(r)$ stands for the skyrmion angle, $M_s$ is the saturation magnetization of the chiral ferromagnet film, and $\bm{e_r}$ and $\bm{e_z}$ are unit vectors along the radial direction and the $z$-axis (perpendicular to the interface), respectively.

The spatial distribution of the vector potential $\bm{A}_{\rm Sk}$ is governed by the Maxwell--London equation:
\begin{gather}
    \nabla \times \left( \nabla \times \bm{A}_{\rm Sk} \right) +  \lambda_L^{-2}\Theta(-z)\Theta(z+d_S)\bm{A}_{\rm Sk}
    \notag \\
     = 4 \pi \color{black} \Theta(z)\Theta(d_F-z) \color{black} \nabla \times \bm{M}_{\rm Sk} ,\label{eq2}
\end{gather}
where \color{black} $\Theta(x)$ denotes the Heaviside step function (with $\Theta(0)=1$) and \color{black} $\lambda_L$ stands for the London penetration depth. The Maxwell--London equation should be supplemented by the boundary conditions of continuity of 
\color{black} the normal component of $\bm{B}_{\rm Sk}=\nabla\times \bm{A}_{\rm Sk}$ and tangential component of $\bm{B}_{\rm Sk} - 4\pi \bm{M}_{\rm Sk}\Theta(z)\Theta(d_F-z)$
\cite{LL8}. \color{black}

Since the right hand side of Eq. \eqref{eq2} is proportional to the unit vector $\bm{e_\varphi}$, the vector potential $\bm{A}_{\rm Sk}$ has only the azimuthal component $A_{{\rm Sk},\varphi}$ that depends on $r$ and $z$. \color{black}
The component $A_{{\rm Sk},\varphi}$ is continuous at $z=-d_S, 0, d_F$; its derivative  $\partial A_{{\rm Sk},\varphi}/\partial z$ is continuous at $z=-d_S$ and has the jumps at $z=0$ and $z=d_F$: $\partial A_{{\rm Sk},\varphi}/\partial z|_{z=-0}^{z=+0}=-4\pi M_{{\rm Sk},r}$ and $\partial A_{{\rm Sk},\varphi}/\partial z|_{z=d_F-0}^{z=d_F+0}=4\pi M_{{\rm Sk},r}$.
\color{black}

The solution for $A_{{\rm Sk},\varphi}(r,z)$ can be cast 
\color{black} as the sum of two terms, $A_{{\rm Sk},\varphi}(r,z)
= A_{{\rm Sk},\varphi}^{(+)}(r,z)+\eta A_{{\rm Sk},\varphi}^{(-)}(r,z)$, where 
\color{black}
\begin{gather}
\color{black} A_{{\rm Sk},\varphi}^{(\sigma)}(r,z) =  - \int\limits_0^\infty dq\, J_1(qr) \frac{G^{(\sigma)}(q)}{q} \hspace{3.5cm}{} \notag \\
\color{black} \times
\begin{cases}
    \varkappa_2^{V,(\sigma)} e^{-qz}, & \quad z\geqslant d_F , \\
\frac{1+\sigma}{2}+ \varkappa_1^{F,(\sigma)} e^{qz} + \varkappa_2^{F,(\sigma)} e^{-qz}, & \quad d_F>z\geqslant 0, \\
 \varkappa_1^{S,(\sigma)} e^{Qz} + \varkappa_2^{S,(\sigma)} e^{-Qz}, & \quad 0> z\geqslant -d_S , \\
 \varkappa_1^{V,(\sigma)} e^{qz}, & \quad -d_S>z .
\end{cases}
\label{eq:sol:ASk}
\end{gather}
Here $J_n(z)$ stands for the Bessel function of the first kind.
Also we introduced $Q=\sqrt{q^2+1/\lambda_L^{2}}$ and the functions
\begin{equation}
\color{black} 
\begin{split}
G^{(+)}(q) & \color{black} = - 4\pi M_s \int\limits_0^\infty dr\, r J_1(qr) \theta^\prime(r) \sin\theta(r) ,  \\
G^{(-)}(q) & = - 4\pi M_s \int\limits_0^\infty dr\, r  q J_1(qr) \sin\theta(r) .    
\end{split}
\label{eq:funG}
\end{equation}
\color{black}
Here and afterwards, we use the following  notation $\theta^\prime(r)\equiv d\theta/dr$. \color{black}
Using the continuity of the azimuthal component of the  vector potential, $A_{{\rm Sk},\varphi}$, and \color{black} the boundary conditions for its \color{black} derivative, $\partial A_{{\rm Sk},\varphi}/\partial z$, at $z=-d_S, 0, d_F$, we obtain ($\sigma=\pm$),
\color{black}
\begin{gather}
\varkappa_2^{V,(\sigma)} = \frac{\sigma }{2}(e^{q d_F}-1) - \frac{\sinh(Q d_S) \mathcal{X}}{q\lambda_L^{2}} , \, \varkappa_1^{V,(\sigma)} = 2 Q e^{q d_s} \mathcal{X} ,
\notag \\
\varkappa_1^{F,(\sigma)} = - \frac{1}{2}e^{-q d_F}, \,\,
\varkappa_2^{F,(\sigma)}=- \frac{\sigma}{2} -  \frac{\sinh(Q d_S) \mathcal{X}}{q\lambda_L^{2}} ,\notag \\
\varkappa_1^{S,(\sigma)} = (Q+q) e^{Q d_s} \mathcal{X}, \quad  \varkappa_2^{S,(\sigma)} = (Q-q) e^{-Q d_s} \mathcal{X},\notag \\
\mathcal{X}=\frac{q (1-e^{-q d_F})}{(Q+q)^2e^{Q d_S}-(Q-q)^2e^{-Q d_S}} .
\end{gather}
\color{black}

The current density in the superconducting film, i.e. at $-d_S\leqslant z\leqslant 0$, can be calculated by means of the London equation, $\bm{j} = - \bm{A}_{\rm Sk}/(4\pi \lambda_L^2)$. 
It is more convenient to trace the total supercurrent flowing in the superconducting film, $J_\varphi(r)=\int_{-d_S}^0 dz j_\varphi(r,z)$. Then, we retrieve  $J_\varphi = J_\varphi^{(+)}+\eta J_\varphi^{(-)}$, where 
\begin{equation}
    J_\varphi^{(\pm)} = 
    \int\limits_0^\infty dq \ \frac{J_1(qr)}{4\pi \lambda_L^2} \frac{G^{(\pm)}(q) (1-e^{-q d_F}) (1-e^{-Qd_S})}{Q[q+Q -(Q-q)e^{-Q d_S}]} ,
    \label{eq:supercurrent}
\end{equation}
We mention that this expression is similar to the expression for the current induced by a domain wall \cite{Burmistrov2005}. In the limit of a thick superconductor, $d_S\gg \lambda_L, R$, Eq. \eqref{eq:supercurrent} transforms into the result of Ref. \cite{Dahir2020}. \color{black} Here $R$ stands for the characteristic spatial scale (radius) of a skyrmion. \color{black}

Below we shall focus on the case of a thin chiral ferromagnet, $d_F\ll R$, and a thin superconducting film, $d_S \ll \lambda_L, R$.  As we shall demonstrate in the next section, the asymptotic behavior of the supercurrent can be found for an arbitrary smooth skyrmion profile with $\theta(0)=\pi$ and $\theta(r\to \infty)\to 0$. Commonly used variational examples with such kind behavior are the exponential ansatz $\theta(r) = \bar{\theta}(r/R)$ where $\bar{\theta}(x)=\pi \exp(-x)$ and the 360-degree domain wall ansatz $\bar{\theta}(x) = 2 \arctan(\sinh(R/\delta)/\sinh(R x/\delta))$. Also we shall consider the linear ansatz with $\theta(r) = \pi(1-r/R)$ for $r<R$ and zero overwise.

\begin{figure*}[t]
\centerline{\includegraphics[width=\textwidth]{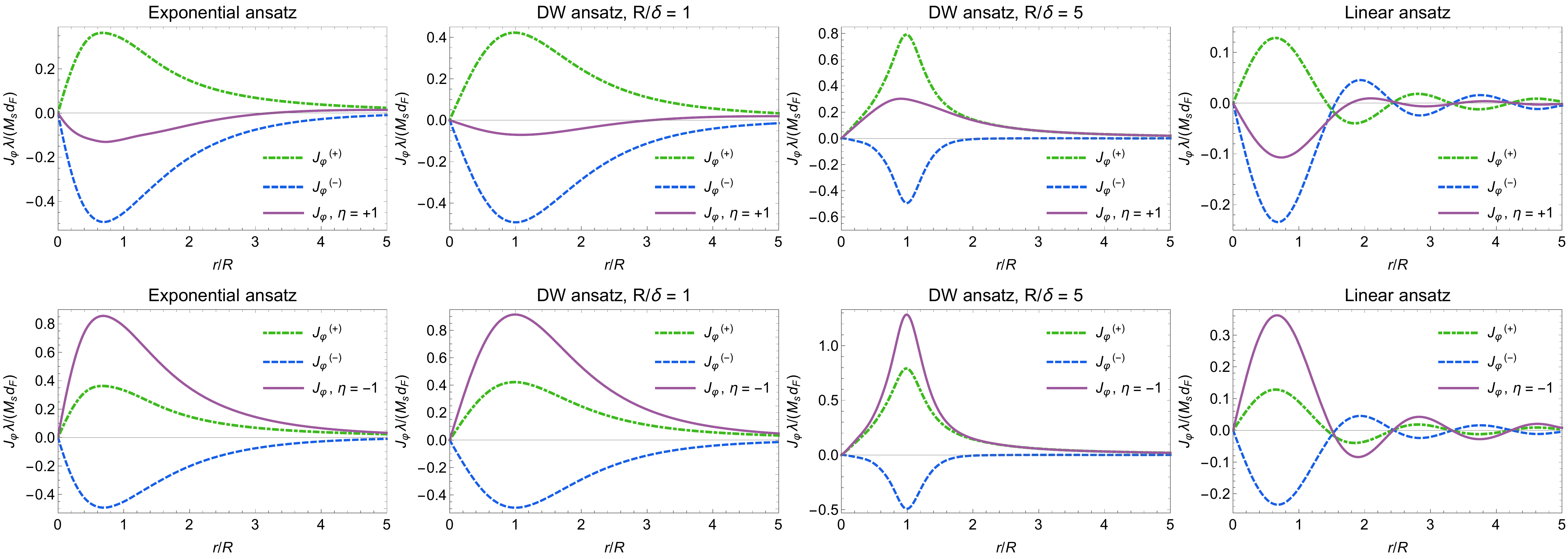}}
    \caption{The dependence of $J_\varphi^{(\pm)}$ and 
        the total supercurrent $J_\varphi$ on  the distance from the skyrmion center for the cases of positive (upper row) and negative (lower raw) chiralities. The parameters are 
        $d_S/\lambda_L=0.01$, $d_F/\lambda_L=0.005$, and $R/\lambda_L=3$. The supercurrent is calculated for the exponential, domain wall (DW) and linear ansatz.}
    \label{fig:Figure3}
\end{figure*}

\subsection{The case of a smooth skyrmion profile}

The behavior of the  supercurrent with the distance from the center of the skyrmion is controlled  by the functions  $G^{(\pm)}(q)$, see Eq. \eqref{eq:funG}. It is convenient to introduce the dimensionless functions  $g^{(\pm)}$,  such that \color{black} $G^{(\pm)}(q) = 4\pi M_s R\ g^{(\pm)}(qR)$, where \color{black} 
\begin{equation}
\begin{split}
     g^{(+)}(y) & =-\int\limits_0^\infty dx \ x J_1(y x) \bar\theta^\prime(x) \sin\bar\theta(x) ,
     \\
     g^{(-)}(y) & =- y \int\limits_0^\infty dx\  x   J_1(y x) \sin\bar{\theta}(x) .
    \end{split}
    \label{eq:def:gq}
\end{equation}
Then in the case of a thin superconducting film, $d_S \ll \lambda_L, R$, and a thin chiral ferromagnet, $d_F\ll R$, Eq. \eqref{eq:supercurrent} can be drastically simplified,
\begin{equation}
    J^{(\pm)}_\varphi(r) = M_s \frac{d_F}{R}
    \int\limits_0^\infty dy \ \frac{y g^{(\pm)}(y) J_1(y r/R) }{1+2 y\lambda/R} .
    \label{eq:supercurrent:thin}
\end{equation}
Here $\lambda=\lambda_L^2/d_S$ denotes the Pearl penetration length \cite{Pearl1964}.
The asymptotic behavior of the function $g^{(+)}(y)$ is given as (see Appendix \ref{App1}),
\begin{equation}
  g^{(+)}(y)= \begin{cases}
      \displaystyle 2 c_2 y , & \quad y\ll 1, \\
      \displaystyle -9 \bar{\theta}^\prime(0)\bar{\theta}^{\prime\prime}(0)/(2 y^4) , & \quad y\gg 1 , 
    \end{cases}  
    \label{eq:g:asymp}
\end{equation}
where we introduced the numerical constants 
\begin{equation}
    c_k = -\frac{1}{4}\int\limits_0^\infty dx\, x^k \bar{\theta}^\prime(x) \sin\bar{\theta}(x), \quad k=-1,0,1, \dots 
\end{equation}
For example, in the case of the exponential ansatz one finds $c_2\approx 0.51$. 
\color{black} The asymptotics  of the function $g^{(-)}(y)$ can be written as (see Appendix \ref{App1}),
\begin{equation}
  g^{(-)}(y)= \begin{cases}
      \displaystyle - b_2 y^2/2 , &  \quad y\ll 1, \\
      \displaystyle  - 3 \bar{\theta}^{\prime \prime}(0) /(2 y^3) , & \quad y\gg 1 . 
    \end{cases}  
    \label{eq:g:asymp:-}
\end{equation}
\color{black}
Here we introduced the numerical constants, 
\begin{equation}
b_k = \int\limits_0^\infty dx\, x^k \sin \bar\theta(x), \quad k=-1,0,1, \dots 
\end{equation}
We note that $b_2\approx 5.94$ in the case of the exponential ansatz.

Let us first consider the case of the skyrmion size much smaller than the size of the vortex, $R\ll \lambda$. Evaluating the integral over $q$ in Eq. \eqref{eq:supercurrent:thin}, we obtain asymptotic behavior of the two components of the supercurrent (see Appendix \ref{App1}),
\begin{equation}
    J_\varphi^{(+)} = \frac{M_s d_F}{\lambda} 
    \begin{cases}
        c_{-1} r/R , & r\ll R ,\\
       c_2 R^2/(2r^2), & R\ll r\ll \lambda, \\
       12 c_2 \lambda^2 R^2/r^4 , & \lambda \ll r ,
    \end{cases}
    \label{eq:current:asym:thin}
\end{equation}
and
\color{black}
\begin{equation}
    J_\varphi^{(-)} = \frac{M_s d_F}{\lambda} 
    \begin{cases}
        \bar{\theta}^\prime(0) r/(2R) , & r\ll R ,\\
      b_2  R^3/(8\lambda r^2)  , & R\ll r\ll \lambda, \\
       3 b_2 \lambda R^3/(2r^4) , & \lambda \ll r .
    \end{cases}
    \label{eq:current:asym:thin:-}
\end{equation}
\color{black}
We note that for $\bar\theta(x) = \pi \exp(-x)$ one finds $c_{-1}\approx 1.17$. 
\color{black} The asymptotic expressions \eqref{eq:current:asym:thin} and \eqref{eq:current:asym:thin:-} suggest nonmonotonous spatial dependence of the both contributions $J_\varphi^{(\pm)}$ to 
the supercurrent with the extremum at the distance of order of the skyrmion radius $R$.
As it is shown in Fig. \ref{fig:Figure3},
the value of $J_\varphi^{(+)}$ ($J_\varphi^{(-)}$) at the extremum  is positive (negative) in the case of exponential and domain wall ansatz. Thus the total supercurrent seems to be sensitive to the skyrmion chirality. In the case of the exponential ansatz the sign of the extremal value of the supercurrent is opposite to the chirality. In the case of the domain wall ansatz the sign of the supercurrent at the extremum depends also on the ratio $R/\delta$.    
\color{black}

In the case of large skyrmion and small Pearl length, $R\gg \lambda$, the part of the supercurrent, $J_\varphi^{(+)}$, which is related with the $z$-component of the skyrmion magnetization, can be found to the lowest order in $\lambda/R$ as (see Appendix \ref{App1}),
\begin{equation}
J_\varphi^{(+)} = - M_s \frac{d_F}{R}\bar\theta^\prime(r/R) \sin \bar\theta(r/R) .
\label{eq:current:asym:thick:1}
\end{equation}
We note that $J_\varphi^{(+)}$ coincides with the current $(\nabla \times \bm{M}_{\rm Sk})_\varphi$ integrated over the width of the chiral ferromagnet.

If the function $\bar\theta(x)$ decays at $x\to \infty$ faster than $1/x^{3}$, the expression \eqref{eq:current:asym:thick:1} determines $J_\varphi^{(+)}$ at $r\ll r_\lambda$ only. 
Then at distances $r\gg r_\lambda\gg R$ the asymptotic behavior of the supercurrent is given as (cf. Eq. \eqref{eq:current:asym:thin}),
\begin{equation}
J_\varphi^{(+)} = 12 c_2 M_s \frac{d_F \lambda R^2}{r^4} , \quad r_\lambda \ll r.
\label{eq:current:asym:thick:2}
\end{equation}
The length scale $r_\lambda$ can be estimated from the condition $|\bar{\theta}(r_\lambda/R)|^2 \sim \lambda R^3/r_\lambda^4$. In the case of the exponential ansatz one finds $r_\lambda \sim R \ln (R/\lambda) \gg R$.

\color{black}
The asymptotic expressions for component $J_\varphi^{(-)}$ of the supercurrent read (see Appendix \ref{App1}),
\begin{equation}
J_\varphi^{(-)} = \frac{3 M_s d_F r}{4R^2} 
    \begin{cases}
       \bar{\theta}^{\prime \prime}(0) \ln(r/R), & \quad r\ll R ,\\
       2b_2 R^5/r^5, &\quad r\gg R . \\
    \end{cases}
    \label{eq:Jphi:-:R1}
\end{equation}
We mention that in the case of $R\gg \lambda$ the dependence of the supercurrent on the distance is qualitatively similar to the case of a skyrmion of a small radius $R\ll \lambda$.  We emphasize that there is a change of the sign of the supercurrent at some distance from the center of the N\'eel--type skyrmion in some cases, see Fig. \ref{fig:Figure3}. Such change of sign can also occurs in the case of a thick superconductor--ferromagnet--superconductor structure \cite{Dahir2020}.
 \color{black}

\subsection{The case of the linear ansatz}

In the case of the linear ansatz the expression \eqref{eq:def:gq} for the function $g^{(+)}(y)$ should be modified in order to have continuous solution for $A_\varphi$ at $r=R$, 
\begin{gather}
g^{(+)}(y) \to g_L^{(+)}(y) = y \int\limits_0^1 dx x J_0(y x) \left [\cos\left(\pi x\right) + \frac{4}{\pi^2}\right ]
\notag \\
\equiv g^{(+)}+\delta g^{(+)} .   
\label{eq:G:linear}
\end{gather}
Here the function $g^{(+)}(y)$ is given by Eq. \eqref{eq:def:gq} and $\delta g^{(+)}(y) = -4 c_2J_1(y)$, where in the case of the linear ansatz, $c_2=1/4-1/\pi^2$. 
Therefore, the function $g^{(+)}_L(y)$ has the following asymptotic behavior,
\begin{equation}
g^{(+)}_L(y)= \begin{cases}
      \displaystyle\frac{\pi^2-6}{2 \pi^4} y^3 , &  \quad y\ll 1, \\
      \displaystyle\frac{\pi^2-4}{\pi^{2}} \frac{\sqrt{2}\cos(y+\pi/4)}{\sqrt{\pi y}} , & \quad y\gg 1. \end{cases}
      \label{eq:G:linear:as}
\end{equation}
We observe that the abrupt change of $\theta(r)$ at $r=R$ results in oscillating behavior of $g^{(+)}(y)$ at $y\gg 1$.

With the help of Eqs.~\eqref{eq:supercurrent:thin}
 and \eqref{eq:G:linear:as}, we obtain the following results for the asymptotic behavior of the supercurrent in the case of $R\ll\lambda$ (see Appendix \ref{App1}),
\begin{equation}
    J_\varphi^{(+)} = \frac{M_s d_F}{4\lambda} 
    \begin{cases}
        (\pi \Si(\pi)-1+4/\pi^2) r/R, & r\ll R ,\\
       3(6-\pi^2) R^4/(\pi^4 r^4), & R\ll r\ll \lambda, \\
       180(6-\pi^2) R^4 \lambda^2/(\pi^4 r^6), & \lambda \ll r .
    \end{cases}
    \label{eq:supercurrent:thin:asym:linear}
\end{equation}
Here $\Si(z)$ stands for the sine integral. We note that in the case of the linear ansatz the $J_\varphi^{(+)}$ component of the supercurrent decays faster at $r\gg R$ than in the case of smooth skyrmion profile. This occurs due to the fact that the contribution to the current from $\delta g^{(+)}(y)$ cancels the leading contributions from $g^{(+)}(y)$. 
As in the case of a smooth skyrmion profile, Eq. \eqref{eq:supercurrent:thin:asym:linear} suggests nonmonotonous behavior of $J_\varphi^{(+)}$ with $r$. There should be the maximum and the minimum in the supercurrent at the distances of the order of the skyrmion size $R$. Contrary to the case of a smooth skyrmion profile, Eq. \eqref{eq:supercurrent:thin:asym:linear} describes asymptotic behavior of the smooth part of $J_\varphi^{(+)}$ only. On the top of the monotononic dependence there is also weak oscillating contribution to $J_\varphi^{(+)}$ with the typical length scale of the order of $R$ as shown in Fig. \ref{fig:Figure3}. This oscillating contribution is the consequence of the abrupt boundary of the skyrmion configuration.

\color{black} The asymptotic behavior of $J_\varphi^{(-)}$ can be read from Eq. \eqref{eq:current:asym:thin:-}. It suggests the existence of the minimum and the maximum at the distance of the order of $R$. Similarly to $J_\varphi^{(+)}$, the contribution $J_\varphi^{(-)}$ has additional oscillations with the distance.
\color{black}

The dependence $J_\varphi^{(+)}(r)$ in the case of large skyrmion size, $R\gg \lambda$, is more intricate. This component of the supercurrent is given as the sum of the contribution discussed above for the case of the smooth skyrmion profile, cf. Eqs. \eqref{eq:current:asym:thick:1} and \eqref{eq:current:asym:thick:2}, and the contribution due to $\delta g^{(+)}(y)$. 
At short distance, $r\ll R$, we find (see Appendix \ref{App1}),
\begin{equation}
J_\varphi^{(+)} = \frac{\pi^2 M_s d_F r}{R^2} \left (1 
 -  3 \frac{\pi^2-4}{\pi^4} \frac{\lambda}{R}\right )
. 
\label{eq:Jphi:linear:large:SK:1}
\end{equation}
In the case of the long distance, $r\gg R$ the contribution to the supercurrent is given as
\begin{equation}
J_\varphi^{(+)} = -  45 \frac{\pi^2-6}{\pi^4} \frac{M_s d_F \lambda R^4}{r^6} .
\end{equation}
We note that in the case of the linear ansatz $J_\varphi^{(+)}$ is stronger suppressed at $r\gg R$ than in the case of a smooth skyrmion profile. The asymptotic behavior of $J_\varphi^{(-)}$ is given by the general expression \eqref{eq:Jphi:-:R1}.

\section{Interaction energy between skyrmion and Pearl vortex\label{Sec:Interaction}}

As above we focus on the case of a thin ($d_S\ll \lambda_L$) superconducting film with a superconducting vortex situated at the distance $a$ from the center of the N\'eel--type skyrmion (see Fig. \ref{fig:Figure1}). In order to compensate the magnetic flux carried by the vortex we assume that there exists anti-vortex located far away from the skyrmion--vortex pair. The free energy of this system, including the magnetic energy of the skyrmion can be written as
\begin{equation}
    \mathcal{F} = \mathcal{F}_{\rm Sk} + \mathcal{F}_{\rm V} + \mathcal{F}_{\rm \overline{V}}+ \mathcal{F}_{\rm Sk-V}+ \mathcal{F}_{\rm Sk-\overline{V}}+\mathcal{F}_{\rm V-\overline{V}} .
    \label{eq:F:gen}
\end{equation}
Here $\mathcal{F}_{\rm Sk}$ denotes the magnetic free energy of the isolated chiral ferromagnet that leads to the appearance of the N\'eel--type skyrmion (see its explicit form in the next section).  
$\mathcal{F}_{\rm V}$ and $\mathcal{F}_{\rm \overline{V}}$ are 
the free energies of the isolated superconducting vortex and anti-vortex, respectively. The electromagnetic interaction between the skyrmion and the vortex is described by the following free energy,
\begin{gather}
    \mathcal{F}_{\rm Sk-V}  = \int \frac{dz d^2 \bm{r}}{4\pi} \Bigl [ \bm{B}_{\rm Sk} \bm{B}_{\rm V}  + \lambda_L^2 (\nabla \times \bm{B}_{\rm Sk}) (\nabla \times \bm{B}_{\rm V})\notag \\
  \times \Theta(-z)\Theta(z+d_S) 
      - 4\pi \bm{M}_{\rm Sk} \bm{B}_{\rm V} \Theta(z)\Theta(d_F-z)
      \Bigr ] 
,
      \label{eq:F-Sk-V-0}
\end{gather}
where $\bm{B}_{\rm V}=\nabla\times \bm{A}_{\rm V}$ and $\bm{B}_{\rm Sk}=\nabla\times \bm{A}_{\rm Sk}$ are the magnetic fields generated by the vortex and the skyrmion, respectively. We note that the first two terms in the right hand side of the expression for $\mathcal{F}_{\rm Sk-V}$ compensate each other in virtue of Eq.
\eqref{eq2}. Therefore, one can have an impression that the distribution of the supercurrent does not influence the interaction energy between the skyrmion and the vortex. In fact, $\mathcal{F}_{\rm Sk-V}$ is intimately related with the supercurrent, see below. In what follows, we shall neglect the free energies of the interaction of the anti-vortex with the skyrmion, $\mathcal{F}_{\rm Sk-\overline{V}}$, and with the vortex, $\mathcal{F}_{\rm V-\overline{V}}$.

The magnetic field of a Pearl vortex in a thin film, $d_S\ll \lambda_L$, can be written in a standard form \cite{Abrikosov-book},
\begin{gather}
{\bm B}_{\rm V} = \phi_0 \sgn(z) \nabla 
\int \frac{d^2\bm{q}}{(2\pi)^2} \frac{e^{-q |z| +i \bm{q}(\bm{r}-\bm{a})}}{q(1+2q\lambda)} .
\end{gather}
Here $\phi_0=h c/2e$ is the flux quantum, $\bm{a}$ is the coordinate vector of the vortex center with respect to the skyrmion center. 
Since $\mathcal{F}_{\rm Sk-V}$ should depend on the distance $a$ between the skyrmion and the vortex only, we can average the magnetic field ${\bm B}_{\rm V}$ over directions of the vector $\bm{a}$. This procedure implies that 
\begin{gather}
\bm{B}_{\rm V} \to - \phi_0 \int\limits_0^\infty \frac{d q}{2\pi} \frac{q\ e^{-q |z|}}{1+2 q\lambda} J_0(qa) \Bigl [ \sgn(z) J_1(qr) \bm{e_r} \notag \\
+ J_0(qr) \bm{e_z}\Bigr ]  .
\label{eq:vortex:field}
\end{gather}
\color{black} We emphasize that the magnetic field $\bm{B}_{\rm V}$ is directed along $-\bm{e}_z$ at the vortex center. The opposite case can be obtained by reversing the sign of the flux quantum $\phi_0 \to -\phi_0$ in expressions below. \color{black}

The free energy of the Pearl vortex (as well as anti-vortex) in a thin superconducting film is given by \cite{Pearl1964}
\begin{equation}
    \mathcal{F}_{\rm V} = \mathcal{F}_{\rm \overline{V}} =  \frac{\phi_0^2}{16 \pi^2 \lambda} \ln \frac{\lambda}{\xi},
\end{equation}
where the superconducting coherence length is assumed to be much shorter than the Pearl length, $\xi\ll\lambda$.

Using Eqs.~\eqref{eq1} and \eqref{eq:vortex:field}, we express the interaction part of the free energy \eqref{eq:F-Sk-V-0} as
\begin{gather} 
\mathcal{F}_{\rm Sk-V} = 
M_s\phi_0 d_F+M_s \phi_0  \int \limits_0^\infty dq \frac{1-e^{-q d_F}}{1+2 q \lambda} J_0(qa)\int\limits_0^\infty dr \, r
\notag \\
\times
 \Bigl [ \eta J_1(qr)\sin\theta(r) 
+J_0(qr) \bigl (\cos\theta(r)-1\bigr ) \Bigr]    .
\label{eq:Int:Sk:V:0}
\end{gather}
We note that the first term in the right hand side of Eq. \eqref{eq:Int:Sk:V:0} corresponds to the homogeneous magnetization of the ferromagnetic film. 
\color{black} 
Using the relation $x J_0(x)=d(x J_1(x))/dx$ and the definition \eqref{eq:funG}, the above expression can be rewritten as 
\begin{gather} 
\mathcal{F}_{\rm Sk-V} 
=
M_s\phi_0 d_F- \frac{\phi_0}{4\pi} \int \limits_0^\infty dq \frac{1-e^{-q d_F}}{q(1+2 q \lambda)} J_0(qa)
\notag \\
\times
\Bigl [G^{(+)}(q)+\eta G^{(-)}(q) \Bigr ].
\label{eq:Int:Sk:V}
\end{gather}
We emphasize that in agreement with general expectations \cite{Abrikosov-book}, the interaction part of the free energy can be expressed in terms of the supercurrent as, $\mathcal{F}_{\rm Sk-V}=M_s\phi_0 d_F - \phi_0\int d^2\bm{r} J_\varphi(r)/(2\pi|\bm{r}-\bm{a}|)$. This implies that the derivative of the free energy  with respect to the vortex position yields the supercurrent \eqref{eq:supercurrent}, $J_\varphi(a) = \phi_0^{-1}(\partial \mathcal{F}_{\rm Sk-V}/\partial a)$, cf. Eqs. \eqref{eq:supercurrent} and \eqref{eq:Int:Sk:V}.  \color{black} Consequently, when the sign of the current $J_\varphi(a)$ is positive (negative), the vortex placed at a distance $a$ tends to move towards (away from) the skyrmion center. 
Therefore, the equilibrium position of the vortex is determined by the zero of the total supercurrent. \color{black} We note that in the case of the linear ansatz the function $G^{(+)}$ in Eq. \eqref{eq:Int:Sk:V} should be modified in accordance with Eq. \eqref{eq:G:linear}.
\color{black}

Below we analyse the general expression \eqref{eq:Int:Sk:V} in the case of a thin ferromagnetic film, $d_F\ll R, \lambda$. 

\begin{figure*}[t]
\centerline{
    \includegraphics[width=\textwidth]{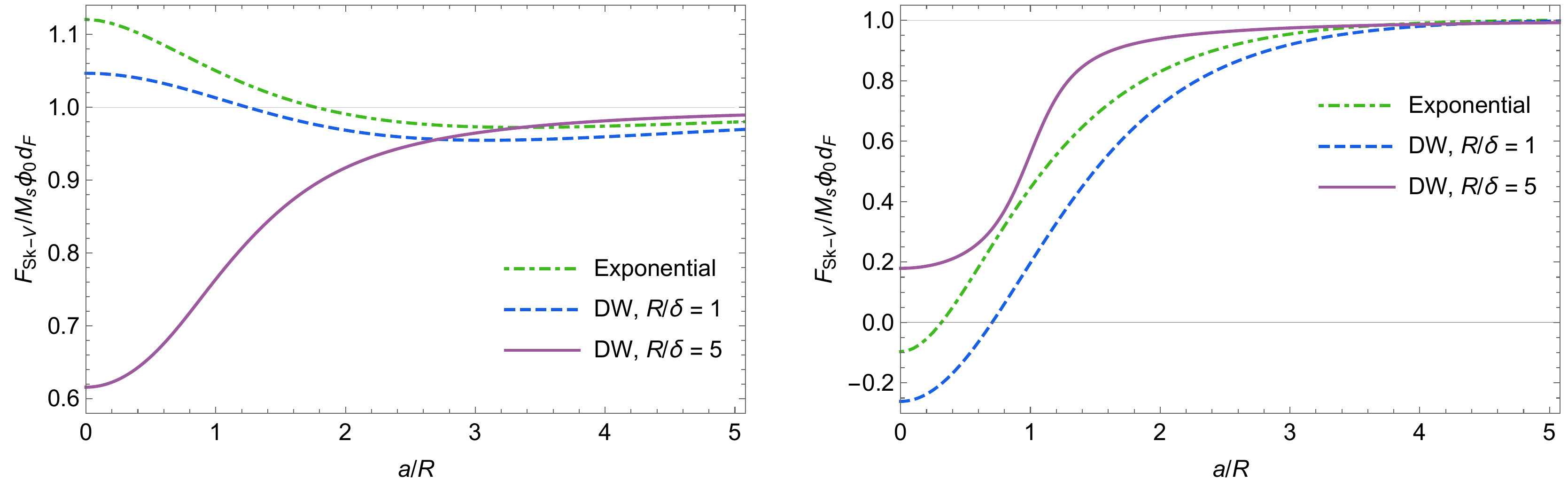}
    }
     \caption{The dependence of the normalized interaction free energy, $\mathcal{F}_{\rm Sk-V}$, on $a/R$ for the chirality $\eta = +1$ (left panel) and $\eta=-1$ (right panel). The ratio of the skyrmion radius and the Pearl length is unity, $\lambda/R = 1$ (see text).}
      \label{fig5}
\end{figure*}


\subsection{The case of a smooth skyrmion profile}

In the case of a smooth skyrmion profile, and for $d_F\ll R, \lambda$,  we find from Eq. \eqref{eq:Int:Sk:V},  
\begin{gather} 
\frac{\mathcal{F}_{\rm Sk-V}}{M_s\phi_0 d_F}=1 + \int \limits_0^\infty dy \frac{J_0(y a/R)}{(1+2 y \lambda/R)} 
\int\limits_0^\infty dx \, x \Bigl [ \eta y  +\bar\theta^\prime(x) \Bigr]  
\notag \\
\times J_1(y x) \sin\bar\theta(x) .
\label{eq:Int:Sk:V:thin}
\end{gather}
As in the case of the supercurrent, we start from the case of a skyrmion of size $R\ll \lambda$. Neglecting unity with respect to $2 y \lambda/R$ in the denominator of the integrand in the right hand side of Eq. \eqref{eq:Int:Sk:V:thin}, we obtain the following asymptotic expression for the interaction free energy at short distances, $a\ll \lambda$, (see Appendix \ref{App2})\footnote{We mention that ${\mathcal{F}_{\rm Sk-V}}/{(M_s\phi_0 d_F)}$ at large distances, $a \gg R$, has a subleading term that depends on chirality, $-\eta b_2 R^3/(8 a\lambda^2)$. This term does not affect the behavior of $\mathcal{F}_{\rm Sk-V}$ with the distance $a$ for the smooth ansatz but becomes essential in the case of the linear ansatz, see Sec. \ref{Sec:lin:F:112}.}
\begin{equation}
\frac{\mathcal{F}_{\rm Sk-V}}{M_s\phi_0 d_F}=1 + \frac{R}{2\lambda} f_\eta\left (\frac{a}{R}\right ) ,
\label{eq:FSkV:asympt:1}
\end{equation}
where the function $f_\eta(z)$ has the following asymptotic behavior 
\begin{gather}
f_\eta(z) = 
\begin{cases} 
\eta b_0 -
4 c_1
+\Bigl(2c_{-1}+\eta \bar\theta^\prime(0)\Bigr ) z^2/2, & \, z\ll 1 ,\\
- 2 c_2/z-c_4/(4z^3), & \, z\gg 1 .
\end{cases}
\label{eq:feta:def} 
\end{gather}

At very long distances, $a\gg \lambda$, the free energy of interaction between the skyrmion and the vortex becomes (see Appendix \ref{App2}),
\begin{equation}
\frac{\mathcal{F}_{\rm Sk-V}}{M_s\phi_0 d_F}=1  - \frac{4 c_2 R^2 \lambda}{a^3} .
\label{eq:FSK-V:long:1}
\end{equation}
We emphasize that at long distances, $a\gg R$, $\mathcal{F}_{\rm Sk-V}$ becomes insensitive to chirality of the N\'eel skyrmion. The coefficient $c_{-1}$ is typically positive whereas $\bar\theta^\prime(0)$ is negative, therefore the interaction free energy may decrease with increase of $a$ for $\eta=+1$. Since the ratio $\mathcal{F}_{\rm Sk-V}/(M_s\phi_0 d_F)$ tends to unity at $a\to \infty$ irrespective of the chirality, one can expect the existence of the minimum of $\mathcal{F}_{\rm Sk-V}$ at some non-zero value of the distance $a$. This situation is realized for the exponential ansatz. In the case of 360-degree domain wall ansatz with $\eta=+1$ 
 the nontrivial minimum exists for $\delta/R \gtrsim 0.64$ only.

Next we consider the opposite case of the skyrmion with the radius much larger than the size of the Pearl vortex, 
$R\gg\lambda$. The interaction free energy can be written as a series in powers of $\lambda/R$ (see Appendix \ref{App2}),
\begin{equation}
    \frac{\mathcal{F}_{\rm Sk-V}}{M_s\phi_0 d_F} = 1 +
    h_{\eta,0}\left (\frac{a}{R}\right )
    +\frac{\lambda}{R} h_{\eta,1}\left (\frac{a}{R}\right ) + \dots
    \label{eq:Fint:Thin}
\end{equation}
The function $h_{\eta,0}$ that determines the magnitude of the interaction free energy has the following asymptotic behavior (see Appendix \ref{App2}),
\begin{equation}
h_{\eta,0}(z) = 
\eta b_{-1}-2 + \left [ \frac{3}{4} \eta \bar\theta^{\prime\prime}(0) \ln z + \bar\theta^{\prime 2}(0)+\eta \beta_0 \right ] \frac{z^2}{2} ,
\label{eq:Fint:Thin:h1s}
\end{equation}
at $z\ll1$, and 
\begin{equation}
h_{\eta,0}(z) = -\frac{\eta b_2}{2 z^3}, \qquad  z\gg 1 .
\label{eq:Fint:Thin:h1l}
\end{equation}
Here the parameter $\beta_0$ is given by the following lengthy expression,
\begin{gather}
\beta_0 = \frac{3}{2} \bar\theta^{\prime}(0)     +\bar\theta^{\prime\prime}(0) \Bigl [\frac{7}{4} - \frac{3(1+2G)}{2\pi}-\frac{6}{\pi}\int\limits_0^1\frac{dx}{x^3}\Bigl (K(x^2)
\notag \\
-\frac{\pi}{2}-\frac{\pi x^2}{8}\Bigr ) \Bigr ]
+ \frac{3}{2} \int\limits_1^\infty dx \frac{\sin\bar\theta(x)}{x^3} 
+\frac{3}{2}\int\limits_0^1 dx \Bigl [\frac{\sin\bar\theta(x)}{x^3}
\notag \\
+\frac{\bar\theta^{\prime}(0)}{x^2}
+ \frac{\bar\theta^{\prime\prime}(0)}{2x}\Bigr ] ,
\label{eq:RlargeL:short}
\end{gather}
where $G\approx 0.916$ denotes the Catalan's constant and $K(x)$ stands for the complete elliptic integral of the first kind. The function $h_{\eta,1}(z)$ that determines the dependence on distance of the subleading contribution to 
$\mathcal{F}_{\rm Sk-V}$ has the following asymptotic behavior (see Appendix \ref{App2}),
\begin{gather}
    h_{\eta,1}(z) = 4 \bigl (2c_{-1}+\eta \bar\theta^{\prime}(0)\bigr )
    +3 \eta \bar\theta^{\prime\prime}(0) z 
    - \Bigl [\frac{9}{4} \bar\theta^{\prime}(0)\bar\theta^{\prime\prime}(0) \ln z
    \notag \\
    + \frac{4}{3}\eta \bigl (\bar\theta^{\prime 3}(0)-\bar\theta^{\prime\prime\prime}(0)\bigr ) -\beta_1 \Bigr ]z^2 , \quad z\ll 1 ,
    \label{eq:Fint:Thin:h2s}
\end{gather}
and
\begin{equation}
    h_{\eta,1}(z) = - \frac{4 c_2}{z^3}, \qquad  z\gg 1 .
\label{eq:Fint:Thin:h2l}
\end{equation}
Here the parameter $\beta_1$ is given as
\begin{gather}
\beta_1 = \frac{9}{2\pi}(1+2G) \bar\theta^{\prime}(0)\bar\theta^{\prime\prime}(0) - \frac{1}{2} \int\limits_{1}^\infty \frac{dx}{x^3}
\partial_x \bigl(x \bar\theta^{\prime}(x) \sin \bar\theta(x)\bigr )
\notag \\
-\frac{1}{2} \int\limits_{0}^1
\frac{dx}{x^3}\partial_x \bigl(x \bar\theta^{\prime}(x) \sin \bar\theta(x)
+ \bar\theta^{\prime 2}(0)x^2+
\frac{3}{2} \bar\theta^{\prime}(0)\bar\theta^{\prime\prime}(0)x^3
\bigr )
\notag \\
+\frac{18}{\pi}
\bar\theta^{\prime}(0)\bar\theta^{\prime\prime}(0)
\int\limits_0^1 \frac{dx}{x^3} \Bigl [K(x^2)-\frac{\pi}{2}-\frac{\pi x^2}{8}
\Bigr]
\notag \\
+\bar\theta^{\prime 2}(0)
-\frac{9}{2} \bar\theta^{\prime}(0)\bar\theta^{\prime\prime}(0) .
\end{gather}

We mention two discrepancies with the case of a skyrmion of a small radius. At first, the short distance behavior of the interaction free energy in the case of $R\gg \lambda$ is not parabolic generically, see Eq. \eqref{eq:Fint:Thin:h1s}. Secondly, the asymptotic behavior of $\mathcal{F}_{\rm Sk-V}$ at $a\gg R$ depends on the skyrmion's chirality. 

\begin{figure*}[t]
    \centering
    \includegraphics[width=\textwidth]{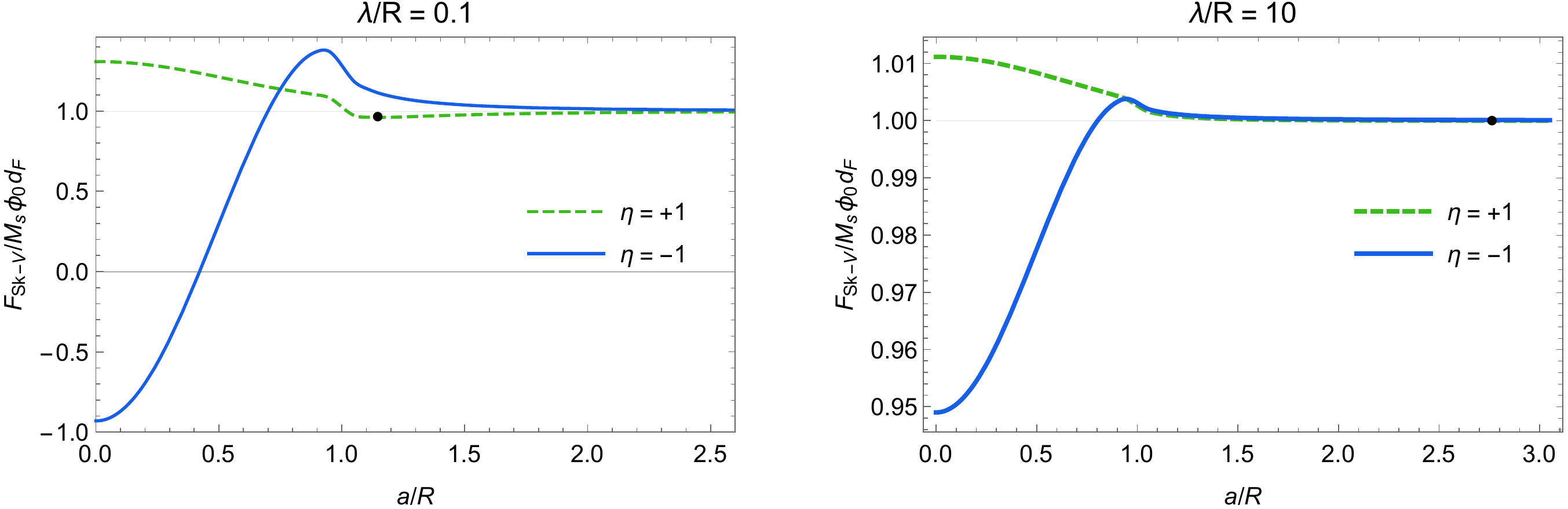}
     \caption{The dependence of the normalized interacting free energy, $\mathcal{F}_{\rm Sk-V}$, on $a/R$ for the linear ansatz for the skyrmion's profile. The plots are for two chiralities and for two values of ratio of the skyrmion radius and the Pearl length: $\lambda/R = 0.1$ (left panel) and $\lambda/R = 10$ (right panel). Black dot near $a \approx 1.2 R$ on the left panel marks the location of the global minimum of the interaction energy. For $\lambda/R = 10$ (right panel) the global minimum is resided at $a \approx 2.8 R$ (see text).} 
    \label{figure:DWa}
\end{figure*}

Provided $\bar\theta^{\prime\prime}(0)>0$, the analytic results \eqref{eq:Fint:Thin:h1s}--\eqref{eq:Fint:Thin:h1l}, suggest the existence of the global minimum of 
$\mathcal{F}_{\rm Sk-V}$ at a certain non-zero distance $a$ in the case of positive skyrmion's chirality $\eta = +1$. 
For negative chirality, $\eta=-1$, the minimum of the interaction free energy is situated at $a=0$. Interestingly, the 360-degree domain wall ansatz is special since $\bar\theta^{\prime\prime}(0)= 0$. Thus, for the 360-degree domain wall ansatz the existence of the minimum in $\mathcal{F}_{\rm Sk-V}$ is controlled by the sign and \color{black} magnitude of $\beta_0$, 
see Eq.\eqref{eq:RlargeL:short}
\color{black} .
For $\delta \gtrsim 0.63 R$ ($\delta \lesssim 0.36 R$) the interaction free energy, $\mathcal{F}_{\rm Sk-V}$, has the minimum at nonzero value of $a$ for the case of positive (negative) chirality, $\eta=+1$ ($\eta=-1$). 

In Figs. \ref{fig5} we show the behavior of the interaction free energy as a function of $a/R$ for both chiralities, $\eta=\pm 1$ and for the skyrmion radius equal to the Pearl length. As one can see, for positive chirality, $\eta=+1$, the minimum of $\mathcal{F}_{\rm Sk-V}$ is reached at nonzero value of the distance $a$.

\color{black} 
We mention that the sign of interacting free energy is determined by the sign of the magnetic flux of the superconducting vortex. If the direction of the magnetic flux at the center of the vortex is opposite to the direction of magnetization at the center of the skyrmion, i.e. magnetic flux is parallel to the vector $\bm{e_z}$, the interacting free energy above will reverse its sign. Then instead of the minimum at $a=0$ (at a finite value of $a$) the minimum will occur at $a=\infty$ (at $a=0$). 
\color{black}

\subsection{The case of the linear ansatz\label{Sec:lin:F:112}}

\color{black} 
As in the case of supercurrent, the interacting free energy for the linear ansatz for the skyrmion profile needs a separate treatment.
The interaction energy 
can be written in the form similar to \eqref{eq:Int:Sk:V},
\begin{gather} 
\mathcal{F}_{\rm Sk-V, \it L} 
=
M_s\phi_0 d_F- M_s\phi_0 R \int \limits_0^\infty dq \frac{1-e^{-q d_F}}{q(1+2 q \lambda)} J_0(qa)
\notag \\
\times
\Bigl [g^{(+)}_L(qR)+\eta g^{(-)}(qR) \Bigr ] .
\label{}
\end{gather}
Here the functions $g^{(-)}$ and  $g^{(+)}_L$ are defined in Eqs.~\eqref{eq:def:gq} and  \eqref{eq:G:linear}, respectively. As it was described in Sec. \ref{Sec:SuperCurrent}, the function $g^{(+)}_L(y)$ is given by a sum of two terms: one identical to the case of the smooth profile $\theta(r)$, $g^{(+)}$, and the other one, $\delta g^{(+)}$, arising due to discontinuity of $\theta^\prime(r)$ at $r=R$, see Eq. \eqref{eq:G:linear}. Accordingly we can represent the free energy as a sum $\mathcal{F}_{\rm Sk-V, \it L} \equiv \mathcal{F}_{\rm Sk-V} + \delta \mathcal{F}_{\rm Sk-V}$. Here $\mathcal{F}_{\rm Sk-V}$ is given by Eq. \eqref{eq:Int:Sk:V:thin} with  $\bar\theta(x) = \pi(1-x)$ for $x \leqslant 1$ and zero overwise. The second term for a thin ferromagnetic film, $d_F \ll R, \lambda$, is defined as
\begin{equation}
    \delta \mathcal{F}_{\rm Sk-V} = - M_s \phi_0 d_F\int \limits_0^\infty dy \frac{J_0(ya/R)}{(1+2 y \lambda/R)} \delta g^{(+)}(y).
    \label{eq:DeltaF:int}
\end{equation}

We have studied the behavior of $\mathcal{F}_{\rm Sk-V}$ in the previous subsection, thus we can focus on examining solely the contribution from $\delta \mathcal{F}_{\rm Sk-V}$. 

Similar to the previous sections we begin with the case of a small skyrmion radius, $R \ll \lambda$. At short distances, $a \ll \lambda$, we present the free energy likewise Eq. \eqref{eq:FSkV:asympt:1},
\begin{equation}
\frac{\delta \mathcal{F}_{\rm Sk-V}}{M_s\phi_0 d_F}= \frac{R}{2\lambda} \delta f_{\eta}\left (\frac{a}{R}\right ) ,
\end{equation}
where $\delta f_{\eta}$ behaves as follows (see Appendix \ref{App2}) 
\begin{gather}
    \delta f_{\eta}(z) = 4 c_2
    \begin{cases} 
    1- z^2/4, & \, z\ll 1 ,\\
    1/(2z)  +1/(16 z^3), & \, z\gg 1 .
    \end{cases}
    \label{eq:asym:Deltaf}
\end{gather}

Collecting both contributions, $\mathcal{F}_{\rm Sk-V}$ and $\delta \mathcal{F}_{\rm Sk-V}$,  together, we can determine the behavior of the free energy $\mathcal{F}_{\rm Sk-V, \it L} \equiv 1+(R/2\lambda) f_{\eta, L}(a/R)$. The function $f_{\eta, L}$ has the following asymptotic behavior at short distances, $z\ll 1$,
\begin{equation}
    f_{\eta, L}(z) = \eta b_0-4(c_1-c_2) +[c_{-1}-c_2+\eta \bar\theta^\prime(0)/2] z^2,
\label{eq:asym:Fint:L1}
\end{equation}
whereas at $z\gg 1$ it becomes
\begin{equation}
    f_{\eta, L}(z) =  (c_2-c_4)/(4z^3) .
    \label{eq:asym:Fint:L2}
\end{equation}
Therefore, at $R\ll r\ll\lambda$ the interacting free energy in the case of the linear ansatz can be written as 
\begin{equation}
    \frac{\mathcal{F}_{\rm Sk-V}}{M_s\phi_0 d_F}= 1+ \frac{(c_2-c_4)R^4}{8 \lambda a^3}- \frac{\eta b_2 R^3}{8a \lambda^2} .
    \label{eq:as:linear:123}
    \end{equation}
Here, also, we add the term of the next order in $R/\lambda$ which depends on the skyrmion chirality (see \cite{Note1}). This term dominates the second term in the right hand side of Eq. \eqref{eq:as:linear:123} for $\sqrt{R\lambda}\ll r \ll \lambda$. Since for the linear ansatz $c_2-c_4=2(\pi^2-6)/\pi^4$, the second term proportional to $1/a^3$ matches with the corresponding asymptotic of the current $J_\varphi^{(+)}$, cf. Eq. \eqref{eq:supercurrent:thin:asym:linear}.

\color{black}

Due to strict localization of the skyrmion and stronger suppression of the supercurrent at distances, $a\gg \lambda$, we expect the interaction energy to decay faster as compared to the case of a smooth profile. Indeed, the expression \eqref{eq:DeltaF:int} yields (see Appendix \ref{App2})
\begin{equation}
    \frac{\delta \mathcal{F}_{\rm Sk-V}}{M_s\phi_0 d_F}= \frac{4c_2 R^2 \lambda}{a^3}, \qquad \lambda \ll a ,
    \label{eq:asym:DeltaF:long}
\end{equation}
that cancels out contribution \eqref{eq:FSK-V:long:1}. Therefore, the interacting free energy at large separations $a$ becomes sensitive to the chirality of skyrmion  opposed to the case of smooth profile, 
\begin{equation}
\frac{\mathcal{F}_{\rm Sk-V}}{M_s\phi_0 d_F}=1  - \frac{\eta b_2 R^3 }{2 a^3} , \qquad \lambda \ll a.
\label{eq:asym:Fint:L3}
\end{equation}

\color{black}
Different asymptotic expressions, Eqs. \eqref{eq:asym:Fint:L1}, \eqref{eq:asym:Fint:L2}, \eqref{eq:as:linear:123}, and \eqref{eq:asym:Fint:L3}, suggest that the vortex resides at a distance $a \sim \sqrt{R \lambda}$ from the center of the skyrmion for $\eta = +1$ and at $a=0$ for $\eta = -1$.
\color{black}

In the opposite case of a large skyrmion radius, $R \gg \lambda$, the additional contribution $\delta \mathcal{F}_{\rm Sk-V}$ to the interacting free energy can be expanded in a series in powers of $\lambda/R$, much the same as Eq. \eqref{eq:Fint:Thin},
\begin{equation}
    \frac{\delta \mathcal{F}_{\rm Sk-V}}{M_s\phi_0 d_F} = \delta h_{\eta,0}\left (\frac{a}{R}\right )
    +\frac{\lambda}{R} \delta h_{\eta,1}\left (\frac{a}{R}\right ) + \dots
    \label{eq:Fint:Thin:L}
\end{equation}

Asymptotic behavior of functions $\delta h_{\eta,0}(z)$, $\delta h_{\eta,1}(z)$ is investigated in Appendix \ref{App2}. Combining them with contributions from $h_{\eta,0}(z)$ and $h_{\eta,1}(z)$ (see Eqs. \eqref{eq:Fint:Thin:h1s}, \eqref{eq:Fint:Thin:h1l}, \eqref{eq:Fint:Thin:h2s}, and \eqref{eq:Fint:Thin:h2l}), we obtain 
\begin{equation}
    h_{\eta,0, L}(z) = \begin{cases} 
        \eta b_{-1}-2 + 4c_2 + \left [\pi^2+\eta \beta_0 \right ] \frac{z^2}{2}, & z \ll 1, \\
        -\eta b_2/(2 z^3), & z \gg 1  ,
    \end{cases}
\end{equation}
and 
\begin{equation}
    h_{\eta,1,L}(z) = 
    8\left(c_{-1} - c_2 -\frac{\pi \eta}{2}\right ) + 6\left (\frac{2\pi^3\eta}{9}-c_2-\frac{\beta_1}{6}\right ) z^2,
\end{equation}
for $z\ll 1$, and 
\begin{equation}
    h_{\eta,1,L}(z) = 
     9 (c_2-c_4)/(2z^5) , \quad z\gg 1 .
     \label{eq:Fint:Thin:h2Linl}
\end{equation}
The above asymptotic expressions suggest that for the positive chirality, $\eta = +1$, the vortex have to be settled at a distance of order $R$ from the skyrmion's center, whereas for $\eta = -1$ the vortex is situated exactly at the center of the skyrmion, $a=0$. 

\color{black} We illustrate the dependence of the interacting free energy on the distance $a$ in the case of the linear ansatz in Fig. \ref{figure:DWa}. On the left panel of Fig. \ref{figure:DWa} one can see the minimum of $\mathcal{F}_{\rm Sk-V}$ (marked by the black dot) for the positive chirality and $\lambda/R=0.1$. For $\lambda/R=10$ (right panel of  Fig. \ref{figure:DWa}) and positive chirality the shallow global minimum of $\mathcal{F}_{\rm Sk-V}$ (also indicated by the black dot) is  located at $a \approx 2.8R$, which is consistent with our prediction, see Eq. \eqref{eq:as:linear:123}. 

\color{black} It should be noted that, in contrast to the case of a smooth ansatz, the transformation $\phi_0 {\to} -\phi_0$ interchanges the qualitative behavior in cases of positive and negative chirality. Namely, for $\eta = +1$ the free energy will have the  minimum 
at $a = 0$ whereas for $\eta = -1$ the minimum of the free energy will be shifted from $a = 0$ to some nonzero $a$. \color{black}

We note that in the case of the linear ansatz the existence of the minimum of the interaction free energy on a finite distance from the skyrmion's center has been noticed in Ref. \cite{Dahir2018}.
\color{black}

\section{The effect of the Pearl vortex on the skyrmion\label{Sec:BackAction}}

The magnetic free energy of the chiral ferromagnetic film  is given by \cite{Bogdanov1989}
\begin{gather}
    \mathcal{F}_{\text{magn}}[\bm{m}] =d_F \int d^2 \bm{r} \bigg\{ A (\nabla \bm{m})^2 + K(1- m_z^2) + \notag \\
    \hspace{1cm} + D \bigl [m_z \nabla \cdot \bm{m} - (\bm{m}\cdot \nabla) m_z \bigr ] \bigg\} .
    \label{eq:MagFe}
\end{gather}
Here $\bm{m}(\bm{r})$ denotes the unit vector of magnetization direction, $A>0$ stands for the exchange constant, $D$ is the Dzyaloshinskii--Moriya interaction, and $K>0$ denotes the perpendicular anisotropy constant. The magnetic free energy is normalized in such a way that $\mathcal{F}_{\text{magn}}$ is zero for the ferromagnetic state, $m_z=1$. \color{black} We note that we include the energy of the magnetic field $\bm{B}_{Sk}$ created by the skyrmion into the definition of the anisotropy constant $K$ (see Appendix \ref{App:DeMag}). \color{black} Substituting  $\bm{m} = \bm{m}_{\rm Sk}=\bm{M}_{\rm Sk}/M_s$, see Eq. \eqref{eq1}, into Eq. \eqref{eq:MagFe}, we find
\begin{gather}
   \mathcal{F}_{\rm Sk} \equiv \mathcal{F}_{\text{magn}}[\bm{m}_{\rm Sk}] = 2\pi d_F \int\limits_0^\infty dr \, r \Biggl\{ A \Bigl[\theta^{\prime 2}(r)+\frac{\sin^2\theta(r)}{r^2}\Bigr ]
   \notag \\+D\eta \Bigl [\theta^\prime(r) +\frac{\sin(2\theta(r))}{2r}\Bigr ]
    + K \sin^2\theta(r)\Biggr \} .
\end{gather}
Assuming a scaling form of the skyrmion profile, $\theta(r)=\bar\theta(r/R)$, we obtain
\begin{equation}
    \mathcal{F}_{\rm Sk} = d_F \Bigl ( \alpha_A A - \alpha_D \eta D R +  \alpha_K K R^2/2 \Bigr ) ,
    \label{eq:Fsk:f}
\end{equation}
where 
\begin{align}
\alpha_A & = 2\pi \int\limits_0^\infty dx\ x     \Bigl[\bar\theta^{\prime 2}(x)+\frac{\sin^2\bar\theta(x)}{x^2}\Bigr ], \notag \\
\alpha_D & = - 2\pi \int\limits_0^\infty dx\ x
\Bigl [\bar\theta^\prime(x) +\frac{\sin(2\bar\theta(x))}{2x}\Bigr ], \notag \\
\alpha_K &= 4\pi \int\limits_0^\infty dx\ x  \sin^2\bar\theta(x) .
\end{align}
We note that $\alpha_{A,D,K}$ are positive constants in the case of the linear and exponential ansatz and are positive functions of the parameter $R/\delta$ in the case of the 360-degree domain wall ansatz. 

\color{black} It is worthwhile to mention that the free energy \eqref{eq:MagFe} does not account for the dipole--dipole interaction. However, since the dipole--dipole energy scales as the first power of the skyrmion radius $R$ (see e.g. \cite{Ezawa2010}), it  can be taken into account by modification of the magnitude of the parameter $\alpha_D$. 
\color{black}

Minimizing  $\mathcal{F}_{\rm Sk}$ with respect to $R$, one can find the optimal radius of the skyrmion 
\begin{equation}
R_0 = \alpha_D|D|/(\alpha_K K)
\end{equation}
 and the chirality $\eta=\sgn D$. We note that the existence of a skyrmion in a chiral ferromagnetic film is possible under  the following condition,
\begin{equation}
    \alpha_A A < \alpha_K K R_0^2/2 .
    \label{eq:cond:SK:0}
\end{equation}


In order to simplify the presentation, we shall start our considerations from the cases of the linear and exponential ansatz. In the presence of vortex anti-vortex pair the skyrmion radius is obtained by minimization of $\mathcal{F}_{\rm Sk}+\mathcal{F}_{\rm Sk-V}$ with respect to $R$ and $a$. Let us start from the case of a skyrmion of small radius, $R_*\ll\lambda$. For the negative chirality the optimal distance between the skyrmion and the vortex is zero. Therefore, as it follows from Eqs. \eqref{eq:FSkV:asympt:1} and \eqref{eq:Fsk:f}, for $\eta=-1$ the interaction between skyrmion and vortex results in increase of the skyrmion radius,
\begin{equation}
R_* = R_0  + (2c_1+b_0/2) \ell_K^2/\lambda .  
\label{eq:R:11}
\end{equation}
Here $\ell_K=\sqrt{M_s\phi_0/(\alpha_K K)}$ is the length scale associated with the anisotropy energy.
In the case of linear ansatz one needs to make the following substitution, $c_1\to c_1-c_2$ in Eq. \eqref{eq:R:11}, see Eq. \eqref{eq:asym:Fint:L1}.

In the case of the positive chirality the optimal distance between the vortex and the skyrmion for the exponential ansatz is proportional to the skyrmion radius, $a_0=\zeta_0 R$, see Eq. \eqref{eq:FSkV:asympt:1}. 
Interestingly, we find that in the case of $\eta=+1$ the skyrmion radius is also enlarged due to interaction with the vortex,
\begin{equation}
R_* = R_0  - f_{+1}(\zeta_0) \ell_K^2/(2 \lambda)  .
\label{eq:R:12}
\end{equation}
We note that $f_{+1}(\zeta_0)<0$. 

\color{black} 
In the case of the linear ansatz with $\eta=+1$, using Eq. \eqref{eq:as:linear:123}, we can find the following result for the skyrmion radius, 
\begin{equation}
R_*=R_0+ \frac{5 b_2^{3/2}}{24 (3(c_2-c_4))^{1/2}}\frac{\ell_K^2 R_*^{3/2}}{\lambda^{5/2}} .
\label{eq:R:12b}
\end{equation}
Although, the above equation predicts ehnancement of the skyrmion radius due to interaction with the vortex, the numerical constant $5 b_2^{3/2}/[24 (3(c_2-c_4))^{1/2}]\approx 0.04$ such that the enhancement is extremely small. The results~\eqref{eq:R:11}--\eqref{eq:R:12b} are applicable for $\lambda\gg\max\{R_0,\ell_K\}$.

In Table \ref{Tab} we present estimates of the change of the skyrmion radius due to interaction with the vortex for several ferromagnet structures. As one can see from the Table \ref{Tab}, the increase of the skyrmion radius $\delta R = R_*-R_0$ is typically small (of the order of a few per cent). Also we note that the estimate of $\delta R$ depends on the form of the skyrmion profile. We mention that the estimates of the bare skyrmion radius $R_0$ given in Table \ref{Tab} on the basis of values of the parameters $D$ and $K$ can significantly deviate from the values actually measured in the experiment.  For example, for the [Ir$_1$Fe$_{0.5}$Co$_{0.5}$Pt$_1$]$^{10}$/MgO/Nb heterostructure the skyrmion radius of the order of 50 nm has been reported \cite{Petrovic2021}. This observation can indicate that in order to estimate $\delta R$ in a realistic structure one needs to find the actual skyrmion profile in the presence of the vortex--anti-vortex pair.
\color{black}

In order a vortex--anti-vortex pair can be spontaneously generated in the presence of a skyrmion the total free energy \eqref{eq:F:gen} should be negative. This implies the following inequality, 
\begin{equation}
    \alpha_A  A - \frac{\alpha_K K R_*^2}{2}+\alpha_K K \ell^2_K + \frac{\phi_0^2}{8\pi^2 \lambda d_F} \ln \frac{\lambda}{\xi} < 0 .
    \label{eq:ineq:12}
\end{equation}
Since $R_*$ is larger than $R_0$ this inequality can be fulfilled provided the condition \eqref{eq:cond:SK:0} holds. We note that then the radius of the skyrmion should satisfy $\lambda\gg R_0\gg \ell_K$. In particular, the vortex--anti-vortex pair cannot be generated spontaneously in the absence of the Dzyaloshinskii--Moriya interaction, i.e. at  $D=0$. Indeed, in the latter case $R_*\ll\ell_K$ and the left hand side of the inequality \eqref{eq:ineq:12} is positive. In fact, there is a minimal value of the Dzyaloshinskii--Moriya interaction at which the spontaneous generation of a vortex--anti-vortex pair is possible,
\begin{gather}
|D| > \left [\frac{2\alpha_K K}{\alpha_D^2} \left (\alpha_A  A +\alpha_K K \ell^2_K + \frac{\phi_0^2 \ln (\lambda/\xi)}{8\pi^2 \lambda d_F} \right )\right ]^{1/2} \notag \\
+ f_\eta(\zeta_0)\frac{\alpha_K K \ell_K^2}{2\alpha_D\lambda}  .
\end{gather}

Now let us assume that the skyrmion radius is large, $R\gg \lambda$. Then, Eqs.~\eqref{eq:Fint:Thin} and \eqref{eq:Fsk:f} result in the following equation for the skyrmion radius modified by the interaction with the vortex,
\begin{equation}
\frac{R_*^3}{R_0^3} - \frac{R_*^2}{R_0^2} = 
h_{\eta,1}(\zeta_0) \frac{\lambda \ell_K^2 }{R_0^3} .
\label{eq:rad:sk}
\end{equation}

For negative chirality, $\eta=-1$, the optimal distance between the skyrmion and the vortex is zero, $\zeta_0=0$. We note that $h_{-1,1}(0)=4 [2c_{-1}-\bar{\theta}^\prime(0)]>0$, see Eq.~\eqref{eq:Fint:Thin:h2s}. For positive chirality, $\eta=+1$, the interaction between skyrmion and vortex has the minimum at finite distance, $\zeta_0\neq 0$. However, as one can check (see Eq. \eqref{b9}), $h_{+1,1}(\zeta_0)>0$. Therefore, for both chiralities 
the skyrmion--vortex interaction leads to increase of the skyrmion radius, 
\begin{equation}
R_* = R_0 \bigl (1+X^{-1/3}+X^{1/3}\bigr )/3,
\label{eq:R*:2}
\end{equation}
where
\begin{equation}
X = 1 + \frac{27 u}{2} + 6\sqrt{3u + 81 u^2}, \quad u = h_{\eta,1}(\zeta_0)\frac{\lambda \ell_K^2}{4R_0^3} .
\end{equation}
We note that for $R_0\ll (\lambda \ell_K^2)^{1/3}$ and $\ell_K\gg \lambda$ the skyrmion radius is parametrically enhanced, $R_*\sim (\lambda \ell_K^2)^{1/3}\gg R_0$.
For $R_0\gg (\lambda \ell_K^2)^{1/3}$, the radius of the skyrmion is only slightly increased, $R_*\sim R_0$. In this case Eq.~\eqref{eq:R*:2} holds under assumption $R_0\gg \lambda$.

A spontaneous generation of the vortex--anti-vortex pair requires the negative total free energy \eqref{eq:F:gen},
\begin{gather}
    \alpha_A  A - \frac{\alpha_K K R_*^2}{2}+\alpha_K K \ell^2_K \left [1+h_{\eta,0}(\zeta_0)+ 2 h_{\eta,1}(\zeta_0)\frac{\lambda}{R_*}\right ] 
    \notag \\
    + \frac{\phi_0^2}{8\pi^2 \lambda d_F} \ln \frac{\lambda}{\xi} < 0 .
\end{gather}
Since $R_* > R_0$ the above inequality can be satisfied provided the condition \eqref{eq:cond:SK:0} holds. However, it can occur only for sufficiently large bare skyrmion radius, $R_0\gg \lambda \gg \ell_K$. In the case $\ell_K\gg R_0\gg \lambda$ the skyrmion radius becomes 
$R_*\sim (\lambda \ell_K^2)^{1/3} \ll \ell_K$. Therefore, the negative term $-\alpha_K K R_*^2/2$ is much smaller than the positive term $\alpha_K K \ell_K^2$ and, consequently, spontaneous generation of vortex--anti-vortex pair is not possible. 

\begin{table*}[t]
\caption{The parameters $M_s$,$A$,$K_u$, and $D$ for a number of thin chiral ferromagnet films. The estimates for the bare radius in zero external field ($R_0$), change of skyrmion radius ($\delta R \equiv R_* - R_0$)  and an anisotropic scale ($\ell_K$) for the exponential ansatz are given. In order to obtain the estimate for change of radius $\delta R$ \color{black}  we choose $\lambda=200$ nm.}
    \begin{tabular}{|l|c|c|c|c|c|c|c|}
\hline
 & PtCoPt \cite{Metaxas2007,Sampaio2013} &  IrCoPt \cite{MoreauLuchaire2016} & PtCoNiCo \cite{Ryu2014} & PdFeIr \cite{Romming2013,Romming2015} & [IrFeCoPt]$^{10}$ \cite{Petrovic2021} \\
\hline
Saturation magnetization $M_s$ ($10^3$ A/m) & 580 & 956 & 600  & 1100 & 1450 \\
\hline
Exchange constant $A$ ($10^{-12}$ J/m) & 15 & 10 & 20  & 2.0 & 13.9 \\
\hline
Anisotropy constant $K_u$ ($10^6$ J/m$^3$) & 0.7 & 0.717 & 0.6 & 2.5 & 1.4\\
\hline
DMI parameter $D$ ($10^{-3}$ J/m$^2$) & +3 & +1.6 & +3 & +3.9 & +2.1\\
\hline
Bare radius $R_0$ ($10^{-9}$ m) & 4.1 & 2.1 & 4.8 & 1.5 & 1.4 \\
\hline
Change of radius $\delta R$ ($10^{-9}$ m) & 0.06 & 0.09 & 0.07 & 0.03 & 0.07 \\
\hline
Anisotropy scale $\ell_K$ ($10^{-9}$ m) & 10 & 13 & 10 & 7.5 & 12 \\
\hline
 \end{tabular}
 \label{Tab}
\end{table*}

In the case of the 360-degree domain wall ansatz Eqs. \eqref{eq:R:11}, \eqref{eq:R:12}, and \eqref{eq:rad:sk} remain valid. However, the value of $\zeta_0$ depends on the ratio $R_*/\delta$. The latter is determined from the minimum of the total free energy with respect to $\delta$. The corresponding analysis  can be performed numerically. As one can check, the following inequalities hold $f_{\eta}(\zeta_0)<0$ and  $h_{\eta,1}(\zeta_0)>0$. These inequalities imply that the skyrmion radius increases always in the presence of a vortex--anti-vortex pair. 
\color{black}

\section{Summary and conclusions\label{Sec:DiscConc}}

To summarize, we have studied an interaction of a N\'eel--type skyrmion and a vortex--anti-vortex pair due to stray fields in a chiral ferromagnet--superconductor heterostructure. We computed the supercurrent in a superconducting film induced by a skyrmion. For thin ferromagnet and superconductor films we found that the supercurrent has the maximum at the distance from the center of a skyrmion that is of the order of the skyrmion radius. It is worthwhile to mention that the supercurrent is sensitive to a profile of the skyrmion and its chirality. For example, in the case of smooth profiles (exponential and domain wall ansatzes), the supercurrent decays monotonously at large distances from the skyrmion center. For the case of a linear profile, there are decaying oscillations of the supercurrent at large distances due to discontinuity in $\theta^\prime(r)$ at $r=R$. Therefore, measurements of dependence of the supercurrent on distance can allow one to extract information on the profile of a skyrmion. We mention that the behavior of the supercurrent with a distance from the center of the skyrmion is qualitatively similar to the behavior of the supercurrent induced in a thin superconducting film by a Bloch domain wall in a ferromagnetic film \cite{Burmistrov2005}. The radius of the skyrmion plays the same role as the width of a domain wall. 

We have also computed the energy of interaction between a N\'eel--type skyrmion and a Pearl vortex. We found that the interaction with a Pearl vortex is sensitive to the skyrmion chirality. In the case of a skyrmion with negative chirality, typically, it is more energetically favourable for a vortex to be attracted to the skyrmion center. This occurs in the cases of linear and exponential skyrmion profiles and for a domain wall ansatz with $\delta\gtrsim 0.36 R$. In the case of positive skyrmion chirality a vortex is situated at a finite distance 
from the center of the skyrmion. This happens for linear and exponential profiles and in the case of domain wall ansatz with $\delta\gtrsim 0.63 R$. For the exponential and domain wall profiles the optimal distance becomes of the order of the skyrmion radius whereas for a linear ansatz the vortex is located at $\max\{R,\sqrt{R\lambda}\}$.

It is worthwhile to mention that in the case of a Bloch--type skyrmion it is always energetically favorable for a vortex to settle at the center of the skyrmion \cite{Dahir2019}. Such a behavior is related with the absence of the radial component of magnetization in a Bloch--type skyrmion. Therefore, the Bloch--type skyrmion interacts with the $z$-component of the magnetic field of a Pearl vortex only. This leads to the absence of terms proportional to the chirality $\eta$ in Eqs. \eqref{eq:feta:def} and   \eqref{eq:Fint:Thin:h1s}. As a result, the function $f_{\eta}(z)$ and $h_{\eta,0}(z)$ behave as increasing parabolas at $z\ll 1$. Such a behavior implies the minimum of the interaction free energy at zero distance between the center of the Bloch-type skyrmion and the Pearl vortex. 

The fact that it is energetically favourable for a Pearl vortex to take place at a finite distance from the center of a N\'eel-type skyrmion might have interesting implications for skyrmion lattices \cite{Balkind2019,Neto2021} and dynamics of skyrmions \cite{Menezes2019} in superconductor--ferromagnet heterostructures \cite{elsewhere}.

We have investigated how a Pearl vortex affects a N\'eel-type skyrmion due to their mutual interaction. We found that a vortex--anti-vortex pair leads to an increase of the radius of the N\'eel--type skyrmion. We note that this result can be contrasted with the case of a Bloch--type skyrmion for which a vortex--anti-vortex pair can either increase or decrease the skyrmion radius \cite{Dahir2019}. 
It is also possible that a vortex--anti-vortex pair will be spontaneously generated in the presence of a N\'eel--type skyrmion provided the skyrmion radius and Pearl penetration length are large enough in comparison with the length associated with the anisotropy energy in a chiral ferromagnet, $\lambda, R_0 \gg \ell_K$.
In the opposite case of small bare skyrmion radius, $R_0\ll \ell_K$, spontaneous generation of a vortex--anti-vortex pair is not possible. \color{black} Although, the relation, $\lambda, R_0 \gg \ell_K$, does not typically holds in chiral ferromagnets (see Table \ref{Tab}), recently, spontaneous generation of vortex--anti-vortex pairs in the [Ir$_1$Fe$_{0.5}$Co$_{0.5}$Pt$_1$]$^{10}$/MgO/Nb heterostructure with N\'eel--type skyrmions of large radius (about 50 nm) and \color{black} positive chirality \footnote{We draw a reader's attention to the fact that in Ref. \cite{Petrovic2021} the geometry of the heterostructure differs from the one considered in our work. In Ref. \cite{Petrovic2021} the ferromagnetic layers are above the superconducting film. 
Our results are applicable for the case of such a geometry provided 
the chirality sign is reversed} has been observed \cite{Petrovic2021}. 
\color{black}

For $R_0\ll (\lambda \ell_K^2)^{1/3}\ll \ell_K$, we predict that a vortex--anti-vortex pair existing in a superconducting film can substantially increase the skyrmion radius: it becomes equal to $R_*\sim (\lambda \ell_K^2)^{1/3}\gg R_0$. The typical values of $R_0$, $\ell_K$, and $R_*$ are listed in Table \ref{Tab}. Abrupt increase of the skyrmion radius can be used as indication of appearance of vortex--anti-vortex pairs in superconducting films.
It is an experimental challenge to detect enhancement of the skyrmion radius in a thin ferromagnet--superconductor heterostructure due to generation of vortex--anti-vortex pair in a superconducting film.

\color{black} Our analysis of the skyrmion stability in the presence of a superconducting vortex was restricted to study of change of the skyrmion radius under assumption that the vortex does not affect the skyrmion profile. In fact, this is not necessary the case and one needs to find the skyrmion profile in the presence of the superconducting vortex from minimization of the total free energy $\mathcal{F}_{\rm Sk}+\mathcal{F}_{\rm Sk-V}$. In particular, we expect that the superconducting vortex can lead to an elongated skyrmion profile \cite{elsewhere}. 
\color{black}

Finally, we mention that it would be interesting to generalize our results to the case \color{black} of skyrmions confined to nanodots \cite{Rohart2013} as well as to \color{black} more exotic magnetic excitations, e.g. antiskyrmions, bimerons, biskyrmions, skyrmioniums, etc. \cite{Tretiakov2021}

\begin{acknowledgements}

The authors are grateful to I. Eremin, Y. Fominov, M. Garst, and A. Petrovi\'c for useful comments. The authors are especially thankful to A. Melnikov for the pointing out the importance of the relation between the supercurrent and the interaction free energy. 
The work was funded in part by Russian Science Foundation under the grant No. 21-42-04410.

\end{acknowledgements}

\appendix

\section{Derivation of the asymptotic expressions for the supercurrent \label{App1}}

In this Appendix we present some details of derivation of asymptotic expressions for the supercurrent.

\subsection{The case of a smooth skyrmion profile}

We start from the case of the smooth skyrmion profile. According to Eq. \eqref{eq:supercurrent:thin} the supercurrent is determined by the functions $g^{(\pm)}(y)$, see Eq. \eqref{eq:def:gq}.

To find the asymptotic behavior of the functions $g^{(\pm)}(y)$ in the case of a small argument, $y \ll 1$, we approximate the Bessel function $J_1(x y)$ by $x y/2$ and find,  
\begin{equation}
\begin{split}
     g^{(+)}(y) & = - y/2 \int\limits_0^\infty dx x^2 \bar{\theta}^{\prime}(x) \sin \bar{\theta}(x) \simeq 2c_2 y,
     \\
     g^{(-)}(y) & =- y^2/2 \int\limits_0^\infty dx\  x^2 \sin\bar{\theta}(x) \simeq -b_2 y^2/2.
    \end{split}
    \label{a1}
\end{equation}

Asymptotic expressions at large arguments, $y\gg 1$, can be found in the following way. Changing the variable $x$ to $x y$ under the integral sign in the definitions of the functions $g^{(\pm)}(y)$, see Eq. \eqref{eq:def:gq}, one can then expand the function $\theta$ in powers of $1/y$. Then, we obtain
\begin{equation}
\begin{split}
     g^{(+)} & = \lim_{\beta \to +0} \int\limits_0^\infty dx J_1(x) e^{-\beta x} \left[ \bar{\theta}^{\prime}(0)+ \frac{3 x}{2y}\bar{\theta}^{\prime \prime}(0) \right] \frac{\bar{\theta}^{\prime}(0) x^2}{y^3} \\
    & \simeq - 9 \bar{\theta}^\prime(0)\bar{\theta}^{\prime\prime}(0)/(2 y^4) ,
     \\
     g^{(-)} & = \lim_{\beta \to +0} y \int\limits_0^\infty dx J_1(x) e^{-\beta x} \left[ \bar{\theta}^{\prime}(0) \frac{x}{y}+ \frac{x^2}{2y^2}\bar{\theta}^{\prime \prime}(0) \right] \frac{x}{y^2}  \\
    & \simeq - 3 \bar{\theta}^{\prime\prime}(0)/(2 y^3).
    \end{split}
    \label{a2}
\end{equation}
Equations \eqref{a1} and \eqref{a2} are equivalent to Eqs. \eqref{eq:g:asymp}, \eqref{eq:g:asymp:-}.

We will now present derivation of asymptotic expressions for the  supercurrent in the case of a small skyrmion, $R \ll \lambda$. At the shortest distances from the center of the skyrmion, one can neglect unity in the denominator of the expressions \eqref{eq:supercurrent:thin} and, then, expand the Bessel function in series in $r/R \ll 1$. Then, we retrieve,

\begin{gather} 
    J^{(\pm)}_\varphi = M_s \frac{d_F r}{4 \lambda R}
    \int\limits_0^\infty dy \; y g^{(\pm)}(y)  .
\label{a3}
\end{gather}
The integral $\int\limits_0^\infty dy \; y g^{(\pm)}(y)$ can be simplified with the help of the following identity $y J_1(x y) = -\partial_x(J_0(x y))$. Then, we obtain
\begin{gather} 
    \int\limits_0^\infty dy \; y g^{(\pm)}(y) = \int\limits_0^\infty dx \; \chi_{0,1}(x) \int\limits_0^\infty dy\; J_0(xy) 
    \notag \\=
    \begin{cases}
    4c_{-1}, & \quad {\rm for\ `+' sign} , \\ 2\bar{\theta}^{\prime}(0), & \quad {\rm for\ `-' sign} ,
    \end{cases} 
\label{a4}
\end{gather}
where $\chi_{0}(x) = [x \sin\bar\theta(x)]^\prime$ and $\chi_{1}(x) = [x \bar\theta^\prime(x) \sin\bar\theta(x) ]^\prime$.
This results in Eqs.~\eqref{eq:current:asym:thin}--\eqref{eq:current:asym:thin:-}.

For the case of long distances, $r \gg R$, we rewrite the expressions for the supercurrent components, $J^{(\pm)}_\varphi(r)$, in a more convenient way, raising the denominator into exponent by means of an additional integration,
\begin{equation} 
    J^{(\pm)}_\varphi = M_s \frac{d_F}{R}
    \int\limits_0^\infty dy \int\limits_0^\infty dt\ e^{-t(1+2 y \lambda/R)} y g^{(\pm)}(y) J_1(y r/R) .
\label{a5}
\end{equation}
Let us first consider the integration with respect to the $y$ variable. Since for $r/R \gg 1$ the integral over $y$ is dominated by small values of $y$, for $g^{(+)}$ we obtain, 
\begin{gather} 
    \int\limits_0^\infty dy \; y e^{-2 y t\lambda/R} J_1(xy) J_1(yr/R) 
    \simeq \frac{x}{2}\int\limits_0^\infty dy \; y^2 e^{-y(2t\lambda/R)} \notag \\\times
    J_1(yr/R) = 
    \frac{x}{2}\frac{3(2\lambda t/r)}{\left[1+(2\lambda t/r)^2\right]^{5/2}} \frac{R^3}{r^3} .
\label{a6}
\end{gather}
and for $g^{(-)}$
\begin{gather} 
    \int\limits_0^\infty dy \; y^2 e^{-2 y t\lambda/R} J_1(xy) J_1(yr/R) 
    \simeq \frac{x}{2}\int\limits_0^\infty dy \; y^3 e^{-y(2t\lambda/R)} \notag \\\times
    J_1(yr/R) = 
    -\frac{3x}{2}\frac{1-4(2\lambda t/r)^2}{\left[1+(2\lambda t/r)^2\right]^{7/2}} \frac{R^4}{r^4} .
\label{a7}
\end{gather}
Hence for the supercurrent we find
\begin{equation}
\begin{split}
     J^{(+)}_\varphi & = 2 c_2  M_s \frac{d_F R^2}{r^3} \int\limits_0^\infty dt \frac{3(2\lambda t/r) e^{-t} }{\left[1+(2\lambda t/r)^2\right]^{5/2}} ,
     \\
     J^{(-)}_\varphi & = \frac{3 b_2}{2}  M_s \frac{d_F R^3}{r^4} \int\limits_0^\infty dt \; e^{-t} \frac{1-4(2\lambda t/r)^2}{\left[1+(2\lambda t/r)^2\right]^{7/2}} .
    \end{split}
    \label{a8}
\end{equation}

In this integral forms for the supercurrent components, $J^{(\pm)}_\varphi$, one can clearly figure out the behavior of $J^{(\pm)}_\varphi(r)$ for $r \ll \lambda$ and $r \gg \lambda$. For $r \ll \lambda$ we can substitute $e^{-t}$ by unity and, then, obtain the asymptotic behaviour at intermediate distances Eqs.~\eqref{eq:current:asym:thin}--\eqref{eq:current:asym:thin:-}, $R \ll r \ll \lambda$. Otherwise, when $r$ is much larger than $\lambda$, we neglect the term proportional to the small parameter $\lambda/r$ in the denominator of \eqref{a8}. Then, one gets the last expressions in Eqs.~\eqref{eq:current:asym:thin}--\eqref{eq:current:asym:thin:-}.

Now we consider the case of large skyrmion $R \gg \lambda$. We start from the limit of short distances $r \ll r_\lambda$. In this regime we neglect the term proportional to $\lambda/R$ in the denominator in the right hand side of Eq. \eqref{eq:supercurrent:thin}. For $J^{(+)}_\varphi$ we can use the identity
\begin{equation}
\int_0^\infty dy \; y J_\alpha(x y) J_\alpha(z y) = \delta(x-z)/x, 
\label{a9}
\end{equation}
and find
\begin{align}
    J^{(+)}_\varphi = & -M_s \frac{d_F}{R} \int\limits_0^\infty dx \bar{\theta}^{\prime}(x) \sin \bar{\theta}(x) \delta(x-r/R) 
\notag\\
    = & -M_s \frac{d_F}{R}\bar\theta^\prime(r/R) \sin \bar\theta(r/R), \;\; r \ll r_\lambda .
\label{a10}    
\end{align} 
Equation \eqref{a10} is equivalent to Eq. \eqref{eq:current:asym:thick:1}.

For the sensitive to chirality component of the supercurrent, $J^{(-)}_\varphi$, the easiest way to derive the asymptotic expression at closest distances is done by employing the relation $J_\varphi(a) = \phi_0^{-1} \partial (\mathcal{F}_{\rm Sk-V}/\partial a)$ and differentiating the expression \eqref{eq:Fint:Thin:h1s} with respect to $z$, inserting $z = r/R$. The derivation of \eqref{eq:Fint:Thin:h1s} is presented in  Appendix \ref{App2}.

The asymptotic expressions \eqref{eq:current:asym:thick:2}--\eqref{eq:Jphi:-:R1} for $r\gg r_\lambda$ can be easily derived from Eq. \eqref{a8}.

\subsection{The case of the linear ansatz}

Let us start from the case of $R \ll \lambda$ and derive the asymptotic expression \eqref{eq:supercurrent:thin:asym:linear} for the non-chiral term in supercurrent, $J^{(+)}$. At shortest distances, $r \ll R$, one can proceed similar to the case of the smooth profile,
\begin{gather} 
    J^{(+)}_\varphi = M_s \frac{d_F}{2 \lambda} \int\limits_0^\infty dy \; g^{(+)}_L(y) J_1(y r/R) .
\label{a11}
\end{gather}
Since $g^{(+)}_L(y) = g^{(+)}(y) + \delta g^{(+)}(y)$, for the first contribution to $J^{(+)}_\varphi$ we can use the expression \eqref{a4}. While for the second term, $\delta g^{(+)}(y) = -4 c_2 J_1(y)$, we apply the identity 
\begin{gather}
    \int\limits_0^\infty dy J_1(y) J_1(yr/R) = \frac{2 R}{\pi r} \left[K(R^2/r^2) - E(R^2/r^2) \right] \notag \\\simeq r/(2R) + O(r^3/R^3), \qquad r \ll R . 
\label{a12}
\end{gather}
Here $K(z)$ and $E(z)$ denotes the complete elliptic integrals of the first and second kinds.
Together, these two contributions give the final result, cf. Eq. \eqref{eq:supercurrent:thin:asym:linear},
\begin{equation} 
    J^{(+)}_\varphi = \left(\pi \Si(\pi)-1+\frac{4}{\pi^2}\right) M_s\frac{d_F r}{4\lambda R}  .
\label{a13}
\end{equation}

To find the behaviour of the $J^{(+)}_\varphi$ at large distances we use the method described near Eq. \eqref{a5} above. The only difference is that instead of the expression for the smooth profile function $g^{(+)}(y)$ we need to use the expression \eqref{eq:G:linear} for $g^{(+)}_L(y)$. 
Then, we retrieve,
\begin{gather} 
    \int\limits_0^\infty dy \, y^2 e^{-2 y t\lambda/R} J_0(xy) J_1(yr/R) \simeq
    - \frac{x^2}{4} \int\limits_0^\infty dy\, y^4 e^{-2 y  t\lambda/R}  \notag \\
\times J_1(yr/R) =
    \frac{x^2 R^5}{4 r^5}\frac{15(2\lambda t/r) \left(4(2\lambda t/r)^2-3 \right)}{\left[1+(2\lambda t/r)^2\right]^{9/2}} .
\label{a14}
\end{gather}
This leads to the following approximate expression,
\begin{gather} 
    J^{(+)}_\varphi = M_s \frac{d_F R^4}{r^5} \frac{6-\pi^2}{2\pi^4} 
 \int\limits_0^\infty dt \; e^{-t} \frac{30 \lambda t \left(4(2\lambda t/r)^2-3 \right)}{r\left[1+(2\lambda t/r)^2\right]^{9/2}} .
\label{a15}
\end{gather}
In the case of $R \ll r \ll \lambda$, the exponent $e^{-t}$ in the right hand side of Eq. \eqref{a15} can be approximated by the unity. Then, we obtain, cf. Eq.~\eqref{eq:supercurrent:thin:asym:linear},
\begin{equation} 
    J^{(+)}_\varphi = - \frac{3(\pi^2-6)}{4\pi^4} M_s\frac{d_F R^4}{\lambda r^4}  .
\label{a16}
\end{equation}
In the limit of longest distances, $r \gg \lambda$, we neglect the terms $(2\lambda t/r)^2$ in the enumerator and denominator under the integral sign in Eq.~\eqref{a15}. Then,  we find,  cf. Eq.~\eqref{eq:supercurrent:thin:asym:linear},
\begin{equation} 
    J^{(+)}_\varphi = - \frac{45(\pi^2-6)}{\pi^4} M_s\frac{d_F\lambda  R^4}{r^6} .
\label{a17}
\end{equation}

Finally, we derive asymptotic expressions for $J^{(+)}_\varphi$ for the case of a large skyrmion radius, $R \gg \lambda$. As it was explained in the main text, we have to combine the contributions from the term $g^{(+)}(y)$, given by Eq. \eqref{eq:def:gq}, and due to  $\delta g^{(+)}(y) = -4c_2 J_1(y)$. Let us start from the limit $r \ll R$. We can use Eq.\eqref{a10} for the asymptotic expression, corresponding to the contribution from $g^{(+)}(y)$. In the case of the linear ansatz it reads $(M_s d_F/R) \pi^2 (r/R)$. In order to find the contribution due to the second term, $\delta g^{(+)}(y)$, we replace the Bessel function $J_1(yr/R)$ by $y r/(2R)$ and expand denominator in powers of $y \lambda/R$. Then, we find
\begin{gather} 
   -2c_2 \frac{r}{R} \lim_{\beta \to +0} \int\limits_0^\infty dy \; e^{-\beta y} y J_1(y) 
\left [ 1- \frac{2 y \lambda}{R} + O\left(\frac{\lambda^2}{R^2}\right )\right ]
\notag\\
    \simeq -12 c_2 \frac{\lambda r}{R^2} .
\label{a18}
\end{gather}
Bringing these two contributions together, we retrieve, cf. Eq. \eqref{eq:Jphi:linear:large:SK:1},
\begin{equation} 
    J^{(+)}_\varphi = \frac{\pi^2 M_s d_F r}{R^2} \left (1 
 -  3 \frac{\pi^2-4}{\pi^4} \frac{\lambda}{R}\right ), \qquad r\ll R .
\label{a19}
\end{equation}

For $r \gg R \gg \lambda$ one can repeat derivation following Eqs.~\eqref{a14} and \eqref{a15}. Then one arrives eventually at the expression \eqref{a17}.

\section{Derivation of the asymptotic expressions for the interaction energy \label{App2}}

\subsection{The case of a smooth skyrmion profile}

In this appendix we present some details of derivation of the asymptotic expressions for $\mathcal{F}_{\rm Sk-V}$.

We start from the case of a small skyrmion and a large vortex, $R \ll \lambda$. In the regime of short distances, $a \ll R$, we can neglect the unity in comparison to the large parameter $\lambda/R$ in the denominator under the integral sign in the right hand side of Eq. \eqref{eq:Int:Sk:V:thin}. Expanding the Bessel function $J_0(y a/R)$ in series of $y a/R$, we obtain
\begin{gather} 
    \frac{\mathcal{F}_{\rm Sk-V}}{M_s\phi_0 d_F}\simeq 1 + \int \limits_0^\infty dy \frac{1-(a/R)^2 y^2/4}{2 y \lambda/R} 
\int\limits_0^\infty dx \, x \Bigl [ \eta y  +\bar\theta^\prime(x) \Bigr]  
\notag \\
\times J_1(y x) \sin\bar\theta(x) .
\label{b1}
\end{gather}
This expression can be easily simplified to the form of Eq. \eqref{eq:FSkV:asympt:1}. 

For the intermediate distances, $R \ll a \ll \lambda$, one can simplify Eq. \eqref{eq:Int:Sk:V:thin}  by using the following identities, 
\begin{align} 
    \int\limits_0^\infty & dy \; J_0(y z) J_1(x y) = \Theta(x-z)/x ,\label{b2} \\
    \int\limits_0^\infty & \frac{dy}{y} \; J_0(y) J_1(x y) = \frac{2}{\pi x} 
\Bigl [    
   E\left(x^2\right) - (1-x^2) K\left(x^2\right) \Bigr ]
   \notag \\
   & \times \Theta(1-x)+ 
   \frac{2}{\pi} E\left(x^{-2}\right) \Theta(x-1) .
\label{b3}
\end{align}
After some simplifications, the expression for the interaction energy can be brought to the form of Eq. \eqref{eq:FSkV:asympt:1}.


The case of the longest distances, $a\gg \lambda$, can be studied in the following way. One can transform the expression in the denominator under the integral sign in the right hand side of Eq.~\eqref{eq:Int:Sk:V:thin} into the exponent with the help of an additional integration, $1/(1+2 y \lambda/R) = \int_0^\infty dt \; e^{-t(1+ 2y\lambda/R)}$. Then expanding the Bessel function $J_1(xy)$ in its argument to the lowest order, we derive Eq.~\eqref{eq:FSK-V:long:1}.

Now let us consider the opposite case of large skyrmion radius, $R \gg \lambda$. Making in Eq. \eqref{eq:Int:Sk:V:thin} expansion in powers of $\lambda/R$, we obtain Eq. \eqref{eq:Fint:Thin} with 
the functions $h_{\eta,0}(z)$ and $h_{\eta,1}(z)$ that are given as
\begin{equation} 
    h_{\eta,0} = \int \limits_0^\infty dx dy \, 
x J_0(y z) J_1(y x) \Bigl [ \eta y  +\bar\theta^\prime(x) \Bigr]
     \sin\bar\theta(x)
\label{b4}
\end{equation}
 and 
\begin{gather} 
    h_{\eta,1} = -2 \int \limits_0^\infty dx dy \, x y J_0(y z) J_1(y x)
\Bigl [ \eta y  +\bar\theta^\prime(x) \Bigr]
\sin\bar\theta(x) .
\label{b5}
\end{gather}

We shall start with the asymptotic behavior of the functions $h_{\eta,0}(z)$ and $h_{\eta,1}(z)$ at $z \ll 1$. Using the following identities $d[x J_1(x y)]/dx = y x J_0(x y)$, $y J_1(x y)=-dJ_0(x y)/dx$, the identity \eqref{a9}, and the relation
\begin{equation}
    \int\limits_0^\infty dy J_0(yz) J_0(yx) =  
    \frac{2}{\pi}\begin{cases}
    K(x^2/z^2)/z , &\quad z \geqslant x ,\\
    K(z^2/x^2)/x, & \quad z < x ,
    \label{b6}
\end{cases}
\end{equation}
we can simplify Eqs. \eqref{b4} and \eqref{b5} as follows
\begin{gather} 
    h_{\eta,0}(z) = \cos \bar\theta(z)-1 + \frac{2\eta}{\pi} \int\limits_0^z \frac{dx}{z} \chi_0(x) K\left(\frac{x^2}{z^2} \right) \notag \\
    + \frac{2\eta}{\pi} \int\limits_z^\infty \frac{dx}{x} \chi_0(x) K\left(\frac{z^2}{x^2} \right) 
\label{b7}
\end{gather}
and
\begin{gather} 
    h_{\eta,1}(z)= - 2 \eta \frac{\chi_0(z)}{z} - \frac{4}{\pi}  \int\limits_0^z \frac{dx}{z} \chi_1(x) 
    K\left(\frac{x^2}{z^2} \right)
    \notag \\
    -\frac{4}{\pi} \int\limits_z^\infty \frac{dx}{x} \chi_1(x) K\left(\frac{z^2}{x^2} \right) ,
\label{b8}
\end{gather}
where $\chi_{0}(x) = [x \sin\bar\theta(x)]^\prime$ and $\chi_{1}(x) = [x \bar\theta^\prime(x) \sin\bar\theta(x) ]^\prime$. Next we rewrite the integrals  over the region $x > z$ in the right hand side of Eqs. \eqref{b7} and \eqref{b8} in the following form,
\begin{gather}
    \frac{2}{\pi} \int\limits_z^\infty \frac{dx}{x} \chi_{0,1} (x) K\left(\frac{z^2}{x^2} \right) =
     \int\limits_z^\infty \frac{dx}{x} \chi_{0,1}(x) + \frac{z^2}{4} \int\limits_z^\infty \frac{dx}{x^3} \chi_{0,1}(x) 
    \notag \\
    + \frac{2}{\pi} \int\limits_z^\infty \frac{dx}{x} \chi_{0,1}(x) \left[  K\left(\frac{z^2}{x^2} \right) - \left(\frac{\pi}{2}+\frac{\pi z^2}{8 x^2}\right) \right]
    .
\label{b9}
\end{gather} 
Written in this way, each of the terms converges at $z\to 0$ and the asymptotic behaviour of $h_{\eta,0}(z)$ and $h_{\eta,1}(z)$ can be easily extracted. Then we reproduce Eqs.~\eqref{eq:Fint:Thin:h1s} and \eqref{eq:Fint:Thin:h2s}.

For large values of the argument, $z \gg 1$, it is enough to consider the term in Eqs. \eqref{b7} and \eqref{b8} which is proportional to the integral over the region $x < z$. 
We can also expand the complete elliptic function of the first kind as  $K\left({x^2}/{z^2} \right) = \pi/2 + \pi x^2/(8 z^2) + 9\pi x^4/(128 z^4) + \dots$. Then one can derive Eqs. \eqref{eq:Fint:Thin:h1l} and \eqref{eq:Fint:Thin:h2l}.

\subsection{The case of the linear ansatz}

For linear ansatz we represent the free energy as a sum $\mathcal{F}_{\rm Sk-V, \it L} \equiv \mathcal{F}_{\rm Sk-V} + \delta \mathcal{F}_{\rm Sk-V}$, see Eq. \eqref{eq:DeltaF:int}. For the analysis of $\mathcal{F}_{\rm Sk-V}$ we refer to the previous subsection, while in this subsection we are examining exclusively $\delta \mathcal{F}_{\rm Sk-V}$ defined in Eq. \eqref{eq:DeltaF:int}. We begin with the case of large vortex and small skyrmion, $\lambda\gg R$. For small distances, $a \ll \lambda$, we neglect the unity in the denominator of the expression under the integral sign in Eq. \eqref{eq:DeltaF:int} and employ the identity \eqref{b3}.
Expanding the expression \eqref{b3} in powers of $x$ and  $1/x$, we obtain Eq. \eqref{eq:asym:Deltaf}. At longest distances one can repeat the derivation, following Eq. \eqref{b3}, eventually arriving at Eq. \eqref{eq:asym:DeltaF:long}.

For large skyrmion radius, $R \gg \lambda$, it is convenient to present $\delta \mathcal{F}_{\rm Sk-V}$ as follows,
\begin{gather}
    \frac{\delta \mathcal{F}_{\rm Sk-V}}{M_s\phi_0 d_F} = 4c_2 \int_0^\infty dy \; J_0(y a/R) J_1(y) \notag\\
     - 8 c_2 \frac{\lambda}{R} \int_0^\infty dy \frac{y J_0(y a/R) J_1(y)}{1+2y \lambda/R}.
\label{b12}
\end{gather}
The first term can be easily simplified using the identity \eqref{b2}. The second term turns into $(\lambda/R)\delta h_{\eta,1}(a/R)$ after setting $\lambda/R \to 0$ in denominator of the integrand. Its asymptotic behavior at small distances, $a \ll R$, can be extracted by expanding $J_0(y a/R)$ in powers of $a/R$,

\begin{equation}
    \lim_{\beta \to +0} \int_0^\infty dy \; e^{-\beta y} y (1-y^2 a^2/(4R^2)) J_1(y) = 1 + \frac{3a^2}{4R^2}.
\label{b13}    
\end{equation}

At large distances, $a \gg R$, the second term in \eqref{b12} is dominated by very small $y$, thus one can substitute $J_1(y)$ with $(y/2 - y^3/16+\dots)$ and find
\begin{gather}
    \lim_{\beta \to +0} \int_0^\infty dy \; e^{-\beta y} y J_0(ya/R) (y/2 - y^3/16) \notag\\
     =-R^3/(2a^3)-9R^5/(16 a^5).
\label{b14}
\end{gather}
The first term cancels out \eqref{eq:Fint:Thin:h2l}, thus, finding the next order in the expansion of the function $h_{\eta, 1}(z)$ (see previous subsection) and summing it up with \eqref{b14}, we find \eqref{eq:Fint:Thin:h2Linl}.

We notice that the second term in \eqref{b12} converges to a discontinuous function in $a = R$ with $\lambda/R \to 0$. However, for all finite values of $\lambda/R$, the function remains continuous.

\color{black}

\section{Magnetic self-energy of the isolated skyrmion \label{App:DeMag}}

The magnetic self-energy of the single isolated skyrmion can be represented as (see e.g. \cite{Dahir2018})
\begin{equation}
    \mathcal{F}_{\rm Sk}^{\rm magn} = -\frac{1}{2} \int\limits_0^{d_F} dz \int d^2 \bm{r} \bm{M}_{\rm Sk} \bm{B}_{\rm Sk} .
\end{equation}
Inserting $\bm{B}_{\rm Sk} = \nabla \times \bm{A}_{\rm Sk}$ and using the exact solution for $\bm{A}_{\rm Sk}$, cf. Eq. \eqref{eq:sol:ASk}, we derive the asymptotic expression for $\mathcal{F}_{\rm Sk}^{\rm magn}$ in the case of a thin ferromagnetic film, $d_F \ll R$, 
\begin{equation}
    \mathcal{F}_{\rm Sk}^{\rm magn} = - 2 \pi d_F M_s^2 \int d^2 \bm{r} (1 - m_z^2) .
\end{equation}

Consequently, contribution from demagnetization field to the total free energy of an isolated skyrmion can be included as a rescaling of the perpendicular anisotropy constant: $K \to K - 2\pi M_s^2$, see Eq. \eqref{eq:MagFe}.

\color{black}

\bibliography{bib-skyrmion}

\end{document}